\documentclass[aps,twocolumn,superscriptaddress,prd, nofootinbib]{revtex4-2}
  
\usepackage[utf8]{inputenc}
\usepackage[T1]{fontenc}
\usepackage{amsmath}
\usepackage{amssymb}
\usepackage{latexsym}
\usepackage{amsfonts}
\usepackage{epsfig}
\usepackage{psfrag}
\usepackage{graphicx}
\usepackage{color}
\usepackage{xcolor}
\usepackage{hyperref}
  
\usepackage{graphicx}

\usepackage{amsmath}
\usepackage{bbm}
\usepackage{bbold}

\usepackage{tikz}
\usepackage{ctable}

\usepackage [autostyle, english = american]{csquotes}
\MakeOuterQuote{"}

\usepackage{ulem}
\usepackage{comment}

\begin{document}

\title{Relativistic Quantum Kinetic Theory: \\Higher order contributions in assisted Schwinger pair production}

\author{James P. Edwards}
\affiliation{Centre for Mathematical Sciences, 
University of Plymouth, Plymouth, PL4 8AA, UK}

\author{Naser Ahmadiniaz}
\affiliation{Helmholtz-Zentrum Dresden-Rossendorf, 
Bautzner Landstra{\ss}e 400, 01328 Dresden, Germany,}
 
\author{Sebastian M. Schmidt}
\affiliation{Helmholtz-Zentrum Dresden-Rossendorf, 
Bautzner Landstra{\ss}e 400, 01328 Dresden, Germany,}
\affiliation{Technische Universität Dresden, 
01062 Dresden, Germany,}

%\author{other contributors}
%\affiliation{somewhere,}

\author{Christian Kohlf\"urst}
\affiliation{Helmholtz-Zentrum Dresden-Rossendorf, 
Bautzner Landstra{\ss}e 400, 01328 Dresden, Germany,}

\date{\today}

\begin{abstract}
Quantum kinetic theory is an important tool for studying non-equilibrium, non-perturbative and non-linear interactions within an open quantum system, and as such is able to provide an unprecedented view on particle production in the relativistic, ultra-high intensity regime of quantum electrodynamics. By re-organising the relativistic quantum transport equations for Abelian plasmas and integrating them with a perturbative expansion, we significantly expand the scope for kinetic theories to further elucidate the peculiarities of particle production at a spectral level. \\
Keywords: Assisted Schwinger effect, Strong-field quantum electrodynamics, relativistic quantum transport, quantum kinetic theory, non-equilibrium quantum many-body physics.
\end{abstract}

\maketitle

\section*{}

\section{Introduction}

Quantum electrodynamics (QED) is widely regarded as the best-tested physical theory conceived \cite{Dirac}. Its success is well-illustrated by its consistency in measurements of value of the fine structure constant, $\alpha = e^{2}/4\pi$ and precision in its prediction of  the electron's anomalous magnetic moment, $g-2$, currently determined perturbatively to 4 loop order \cite{Laporta4}.
%(and numerically to 5 loop order \cite{Volkov5, Kinoshita5}). 
Still, since QED was first developed it has undergone a shift in focus regarding phenomenology, from measuring perturbative single-photon absorption and emission at its inception \cite{Tomonaga, PhysRev.74.1439, PhysRev.75.651, PhysRev.76.790, PhysRev.75.1736, PhysRev.75.486, PhysRev.76.749, PhysRev.80.440}, to modern interpretations in cavity \cite{Raimond}, circuit \cite{RevModPhys.72.545} and strong-field laser-matter interactions \cite{DiPiazza:2011tq, Gonoskov:2021hwf, Marklund:2006my}, testing effects such as vacuum polarisation \cite{PhysRevLett.78.424}, vacuum non-linearities \cite{Ahmadiniaz:2024xob, Turcu:2016dxm} or vacuum decay \cite{Abramowicz:2021zja, PhysRevAccelBeams.22.101301}. Nevertheless, despite its success and wide applicability many areas of the theory remain unexplored, cf. recent reviews \cite{Gelis:2015kya, Fedotov, RUFFINI20101}. In particular, research on out-of-equilibrium nonlinear quantum many-body systems remains in its infancy \cite{PhysRevLett.109.110401, doi:10.1126/science.abi8627, Bocquillon}. Accordingly, a variety of techniques are used to study 
%specific features of
one of the theory's most prominent predictions: particle creation \cite{HeisenbergEuler, Weisskopf, Schwinger:1951nm, Sauter:1931, DiracSea}. These include semi-classical techniques and numerical simulations of various complexity, cf. worldlines methods \cite{PhysRevD.96.076002, PhysRevD.72.105004, PhysRevD.73.065028, PhysRevD.72.065001}, WKB analysis \cite{Kohlfurst:2021skr, Taya:2020dco, Dumlu:2011rr, Strobel:2013vza}, Bogoliubov theory \cite{PhysRevD.83.065028, PhysRevD.79.065027, Smolyansky:1997fcC, Kluger}, lattice theory \cite{PhysRevD.87.105006, PhysRevLett.111.201601} and scattering approaches \cite{Brezin:1970xf, FradkinGitmanI, PhysRevD.53.7162, PhysRevD.101.096009, Heinzl:2010vg, Blackburn:2021cuq, PhysRevA.105.013105}.

In this article we are interested in the interplay between perturbative and non-perturbative QED at the mean-field level. The physical phenomenon under consideration is the (assisted) Schwinger effect \cite{PhysRevLett.101.130404}, or the creation of electron-positron pairs by exposing the quantum vacuum to strong electromagnetic fields. Here we will consider this process occurring through a combination of non-perturbative quantum tunneling \cite{Schwinger:1951nm} produced by a strong background, combined with photon absorption from a secondary source \cite{BreitWheeler, Reiss}. 
\footnote{
A brief note on terminology. In this paper, the term "Schwinger effect" will be used to refer to particle pair production via non-perturbative tunnelling -- a form of vacuum decay. High-frequency pair production will be understood as energy absorption by  massive particles in well-defined quanta. This underscores that we work within mean-field theory, and therefore do not have quantized "photonic" degrees of freedom.
}

The rationale behind this approach is that instead of a strong field doing the entire work to create a particle pair  \cite{Ringwald:2001ib, PhysRevLett.101.200403}, it is known that the absorption of an additional photon from the secondary source serves to reduce the threshold value \cite{PhysRevD.80.111301, PhysRevD.100.116018, Schneider:2014mla, PhysRevD.99.056006, Linder:2015vta, Schneider:2016vrl, PhysRevD.97.096004}, see also the Franz-Keldysh effect in solids as its non-relativistic counterpart \cite{Franz, FKeldysh}. In contrast, this article presents an examination of the distinctive signatures of higher-order effects in a weak-field expansion with respect to the secondary source, see also \cite{Torgrimsson:2016ant, Torgrimsson:2017pzs, Aleksandrov:2018uqb}, but with particular focus on the particle spectrum \cite{Orthaber:2011cm, Fey:2011if, Otto:2018jbs}.

We adopt an open quantum system approach, enabling the application of kinetic theories \cite{PhysRevA.94.052111}. In the context of relativistic, adiabatic QED an initial vacuum state is driven out of equilibrium by an external, time-dependent electric background field to produce a particle pair \cite{Vasak, PhysRevD.105.076017, Zhuang, Carruthers, PhysRevD.44.1825}. The formalism is highly general (unifying Vlasov and Bargmann–Michel–Telegdi equations into a single set of coupled differential equations non-linear in $\hbar$), thereby enabling, \textit{i.a.}, the calculation of the momentum spectrum of the produced particles irrespective of the formation mechanism \cite{Hidaka:2022dmn, PhysRevD.57.6525, Ochs}. The challenge thus far has been to distinguish between different contributions to pair creation, such as two- and four-photon effects or mutual assistance, given that this formalism reveals the entire particle spectrum at once. The aim of distinguishing between processes of different photon multiplicity aligns well with recent complementary work in the context of vacuum polarisation in \cite{Macleod:2023asi}, where contributions associated to absorption of fixed numbers of photons from an intense laser were identified.

Here, we resolve this issue for  pair creation by separating the vector gauge potential into two parts which allows the separation of the underlying transport equations into two sets of coupled equations without losing the non-linear aspects of the theory. In this context, the procedure to obtain a perturbative expansion with respect to the secondary source is demonstrated. 

This approach shares some similarities with the Furry picture of strong field QED \cite{Furry:1951zz}, where the photon field is split into a fixed semi-classical background plus a quantum fluctuation. The former is treated non-perturbatively to define dressed fermion states, while the latter is treated in the usual perturbation theory. Here, we treat the primary part of the background field non-perturbatively and expand in the secondary field. This leads naturally to an expansion in a suitable  nonlinearity parameter, $\xi = e \varepsilon E_{\rm cr.}/m\Omega$, where $\varepsilon E_{\rm cr.}$ is the magnitude of the weak electric field in terms of the critical Schwinger field strength \mbox{$E_{\rm cr.} = m^2/e \approx 1.3 \times 10^{18}$~V/m} and $\Omega$ its characteristic frequency. This ``classical'' non-linearity parameter characterises the likelihood of multiple interactions with photons from the weak field, well-illustrated by the exponential suppression of the pair creation amplitude for electric fields below the Schwinger field strength.

The structure of this paper is as follows. In Sec.~\ref{sec:form} we present the relativistic quantum kinetic equations for linearly polarised, purely time-dependent electromagnetic fields in the mean-field approximation. Then we present their decomposition into strong field contributions with weak field perturbations, which allows them to be separated into two sets, and a perturbative expansion to be developed. In Sec.~\ref{sec:results}, we then solve the corresponding differential equations numerically and discuss their significance for specific examples. A summary of our findings is provided in Sec.~\ref{sec:conc}. Non-essential technical details and background information are included in a supplementary file. Throughout the manuscript we use natural units, $\hbar = c = 1$ and express results in terms of the electron mass, $m = 511$ keV.
   
\section{Theoretical Formalism}
\label{sec:form}

Kinetic theories are widely employed to describe transport phenomena for systems in and out-of equilibrium. Their quantum mechanical counterparts \cite{PhysRev.40.749}, quantum kinetic field theories, extend their applicability into the high-energy regime \cite{PhysRevD.37.2878, PhysRevD.50.2848, PhysRevD.110.L111903}. 

Nevertheless, at the lowest order and in particular for Abelian plasmas, the Boltzmann transport equation, cf. \cite{Seipt:2023bcw}, can be replaced by simpler equations of Vlasov type \cite{Brodin, PhysRevE.60.4725, PhysRevE.107.035204}. Furthermore, we assume that backreaction can be ignored \cite{PhysRevD.40.456}, meaning that the dynamical field created by the charge carriers is negligible  compared to the (essentially fixed) background field driving the time evolution. Finally, in homogeneous quantum systems, spatial diffusion is absent.

For the system under consideration energy is continually deposited by the environment, in this case, a time-dependent electric field. We take the energy density of the field to be sufficiently high such that the particle number of the plasma is not a conserved quantity, as particles can be created dynamically over time (c.f. Schwinger pair creation).

We begin by defining the quantities under consideration (see the Supplemental for more details). We introduce a scalar function representing particle density at fixed momentum, $f(t, \mathbf{p})$. This particle distribution function satisfies the quantum kinetic equation \cite{Best, PhysRevLett.87.193902, PhysRevLett.89.153901, PhysRevLett.102.150404, PhysRevD.108.036019}:
\begin{equation}
 \dot{f} (t, \mathbf{p}) + e \mathbf{E} (t) \cdot \nabla_p f(t, \mathbf{p}) = \mathcal Q[f] (t, \mathbf{p}).
 \label{eq:Vlasov}
\end{equation}
The operator $\mathcal Q$ acts, \textit{i.a.}, as a source term for non-perturbative particle pair creation. Specifically, for linearly polarized external fields $\mathbf{E}(t) = (E(t),0,0)$ it is found that (see Supplemental)
\begin{alignat}{8}
 &\mathcal Q[f] (t, \mathbf{p}) && = Q(t, \mathbf{p}) \, \mathcal{W}[{\rm cos}](t, t_{\rm i}, \mathbf{p}) = \label{eq:Q} \\
 & && \hspace{-1.cm} Q(t, \mathbf{p}) \int_{t_{\rm i}}^t {\rm d}t'\, Q(t', \mathbf{p}) \left( 1 - f(t', \mathbf{p})\right) \cos \left(2 \Theta (t, t', \mathbf{p}) \right), \notag
\end{alignat}
with field-dependent quantities $Q$ and $\Theta$, and the quantum statistical factor $\left( 1 - f(t', \mathbf{p})\right)$ which facilitates, for example, Pauli blocking. Note that in this formalism the fermion dynamics are non-local in time, relying on the entirety of the quantum system's history. Therefore, Eq.~\eqref{eq:Q} represents a non-Markovian problem, which are notoriously challenging \cite{Kluger:1998bm, PhysRevD.59.094005}. In this regard, a physical interpretation of the particle distribution's non-trivial time evolution has been proposed only recently \cite{Ilderton:2021zej, Diez:2022ywi}. 
 
For the purposes of this article we first bring the quantum kinetic equation \eqref{eq:Vlasov} into a more appropriate form. On the basis of the auxiliary quantities $\mathcal{G}=\mathcal{W}[{\rm cos}](t, t_{\rm i}, \mathbf{p})$ and $\mathcal{H}=\mathcal{W}[{\rm sin}](t, t_{\rm i}, \mathbf{p})$, cf. Eq.~\eqref{eq:Q}, Eq.~\eqref{eq:Vlasov} is transformed into a closed set of equations
\begin{alignat}{8}
    & \dot{f} && + e \mathbf{E} \cdot \nabla_p f && = Q \, \mathcal{G}, && && && && && \\  
    & \dot{\mathcal {G}} && + e \mathbf{E} \cdot \nabla_p \mathcal{G} && = Q \, \left( 1- f \right) && - 2 \omega \mathcal{H} && && && &&, \notag \\   
    & \dot{\mathcal {H}} && + e \mathbf{E} \cdot \nabla_p \mathcal{H} && = && + 2 \omega \mathcal{G} && && && &&, \notag
  \end{alignat}
with the one-particle on-shell energy $\omega^2 = m^2 + \mathbf{p}^2$. %We introduce \ck{mass}, $\mathbbm{s}$, and \ck{current} densities, $\mathbbm{v}_\parallel$ and $\mathbbm{v}_\perp$, where the latter represent distributions in directions parallel and perpendicular to the polarization of the electric field, respectively. 
In anticipation of a perturbative expansion, we introduce new functions, $\mathbbm{s}$, $\mathbbm{v}_\parallel$ and $\mathbbm{v}_\perp$. In this way, for example the particle distribution is decomposed into 
\begin{equation}
 \omega f(t, \mathbf{p}) = m \mathbbm{s} + p_\parallel \mathbbm{v}_\parallel + p_\perp \mathbbm{v}_\perp.
 \label{eq:f}
\end{equation} 
Similar expressions hold for $\mathcal {G}$ and $\mathcal {H}$. If instead formulated in terms of canonical momenta, $\mathbf{q} = \mathbf{p} + e \mathbf{A}$, we ultimately find the coupled equations
\begin{alignat}{8}
    & \dot{\mathbbm{s}} && && -2 q_\parallel \mathbbm{v}_\perp && && +2 q_\perp \mathbbm{v}_\parallel && + 2e A \mathbbm{v}_\perp && &&= 0, \label{eq:w1} \\  
    & \dot{\mathbbm{v}}_\parallel && && && && -2 q_\perp \mathbbm{s} && &&  &&= -2m\mathbbm{v}_\perp , \notag \\   
    & \dot{\mathbbm{v}}_\perp && && +2 q_\parallel \mathbbm{s} && && && -2e A \mathbbm{s} && &&= +2m\mathbbm{v}_\parallel. \notag
  \end{alignat} 
  
\subsection*{Developing an expansion scheme}

We start by decomposing the so-far arbitrary gauge potential $\mathbf{A} = \boldsymbol{\mathcal{A}} + \mathbf{A}^{\gamma}$ into two vector potentials, corresponding to a spatially homogeneous pulse, with field strength $\mathcal{E}$ and pulse duration $\mathcal{T}$, with temporal profile
\begin{equation}
  \mathcal{A}(t) = \mathcal{E} E_{\rm cr.} \, t \, \exp(-t^2/\mathcal{T}^2), \label{eq:A1} 
\end{equation}
and a secondary, oscillating field, also homogeneous with field strength $\varepsilon$, pulse envelope $\tau$, frequency $\Omega$, and profile
\begin{equation}
 A^{\gamma}(t) = \varepsilon\, E_{\rm cr.} /\Omega \,  \exp(-t^2/\tau^2) \ \sin(\Omega t). \label{eq:A2}
\end{equation}
(In our gauge the temporal components of the gauge potentials vanish).
% We use a temporal gauge.

The Ansatz that allows us to split the complete phase-space information into components is as follows:
%\begin{equation}
% f = f^{(0)} + f^{(1)} + \sum_i \xi^i \ f^{(2,i)}, \ \text{with} \ \xi = \varepsilon m / \Omega.   
%\end{equation}
\begin{equation}
 f = f^{(0)} + f^{(1)} + \sum_{i=1} \, f^{(2,i)}.  
\end{equation}
This expansion at the probability level holds for all constituents in the decomposition \eqref{eq:f}. For clarity, $f^{(0)}$ is associated with the phase-space distribution for free fermions, $f^{(1)}$ with a non-perturbative "dressing" by the first field, $\mathcal{A}$, and the remaining sum over the $f^{(2,i)}$ provides an order-by-order decomposition of the additional contributions from the presence of the oscillating field, $A^{\gamma}$ (that is, $f^{(2, i)}$ and its components in the decomposition (\ref{eq:f}) are $\mathcal{O}(\xi^{i})$). This proposed perturbation scheme is applicable as long as the Keldysh parameter $\gamma = 1/\xi$ of the oscillating field is larger than one \cite{Keldysh:1965ojf}. 

We proceed order by order in $\xi$, recalling that $A^{\gamma}$ is $\mathcal{O}(\xi)$. At zeroth order we find the algebraic relations 
\begin{align}
q_\parallel \mathbbm{v}_\perp^{(0)} = q_\perp \mathbbm{v}_\parallel^{(0)},\ m \mathbbm{v}_\perp^{(0)} = q_\perp \mathbbm{s}^{(0)},\  q_\parallel \mathbbm{s}^{(0)} = m \mathbbm{v}_\parallel^{(0)}.
\end{align}
We require that at the vacuum level we uniformly have no particle excitations. We find initial values, cf Eq.~\eqref{eq:f},
\begin{equation}
 \mathbbm{s}^{(0)} = -m/\omega ,\ \mathbbm{v}_\parallel^{(0)} = -q_\parallel/\omega ,\ \mathbbm{v}_\perp^{(0)} = -q_\perp/\omega.    
\end{equation}
These solutions act as source terms for the components of $f^{(1)}$, which must satisfy
\begin{alignat}{8}
    & \dot{\mathbbm{s}}^{(1)} \! && \! \! && -2 q_\parallel \mathbbm{v}_\perp^{(1)} \! \! && \! && +2 q_\perp \mathbbm{v}_\parallel^{(1)} \! \! && + 2e \mathcal{A} \mathbbm{v}_\perp^{(0)} + 2e \mathcal{A} \mathbbm{v}_\perp^{(1)} \! \! && \! && = 0, \\  
    & \dot{\mathbbm{v}}_\parallel^{(1)} \! && \! \! && \! \! && \! \! && -2 q_\perp \mathbbm{s}^{(1)} \! \! && \! \! &&  \! \! &&= -2m\mathbbm{v}_\perp^{(1)} , \notag \\   
    & \dot{\mathbbm{v}}_\perp^{(1)} \! && \! \! && +2 q_\parallel \mathbbm{s}^{(1)} \! \! && \! \! && \! \! && -2e \mathcal{A} \mathbbm{s}^{(0)} -2e \mathcal{A} \mathbbm{s}^{(1)} \! \! && \! \! &&= +2m\mathbbm{v}_\parallel^{(1)}. \notag
\end{alignat} 

Correspondingly, the equations at $\mathcal{O}(\xi)$ receive contributions from both lower-order results acting as sources:
\begin{alignat}{8}
    & \dot{\mathbbm{s}}^{(2,1)} && && -2 q_\parallel \mathbbm{v}_\perp^{(2,1)} && && +2 q_\perp \mathbbm{v}_\parallel^{(2,1)} \\
    & && && && && \hspace{-2.0cm} + 2e A^{\gamma} \mathbbm{v}_\perp^{(0)} + 2e \mathcal{A} \mathbbm{v}_\perp^{(2,1)} + 2e A^{\gamma} \mathbbm{v}_\perp^{(1)} && = 0, \notag \\  
    & \dot{\mathbbm{v}}_\parallel^{(2,1)} && && && && -2 q_\perp \mathbbm{s}^{(2,1)} && = -2m\mathbbm{v}_\perp^{(2,1)} , \notag \\   
    & \dot{\mathbbm{v}}_\perp^{(2,1)} && && +2 q_\parallel \mathbbm{s}^{(2,1)} && && && \notag \\
    & && && && && \hspace{-2.0cm} - 2e A^{\gamma} \mathbbm{s}^{(0)} - 2e \mathcal{A} \mathbbm{s}^{(2,1)} - 2e A^{\gamma} \mathbbm{s}^{(1)} &&= +2m\mathbbm{v}_\parallel^{(2,1)}. \notag
  \end{alignat} 
All higher orders in the expansion follow the same routine, thus we use index notation where $i \ge 2$,
\begin{alignat}{8}
    & \dot{\mathbbm{s}}^{(2,i)} && && -2 q_\parallel \mathbbm{v}_\perp^{(2,i)} && && +2 q_\perp \mathbbm{v}_\parallel^{(2,i)} \\ 
    & && && && && \hspace{-1.cm} + 2e \mathcal{A} \mathbbm{v}_\perp^{(2,i)} + 2e A^{\gamma} \mathbbm{v}_\perp^{(2,i-1)} &&= 0, \notag \\ 
    & \dot{\mathbbm{v}}_\parallel^{(2,i)} && && && && -2 q_\perp \mathbbm{s}^{(2,i)} &&= -2m\mathbbm{v}_\perp^{(2,i)} , \notag \\ 
    & \dot{\mathbbm{v}}_\perp^{(2,i)} && && +2 q_\parallel \mathbbm{s}^{(2,i)} \notag \\
    & && && && && \hspace{-1.cm} - 2e \mathcal{A} \mathbbm{s}^{(2,i)} - 2e A^{\gamma} \mathbbm{s}^{(2,i-1)} &&= +2m\mathbbm{v}_\parallel^{(2,i)}. \notag
  \end{alignat} 
By truncating the last, infinite set of equations at a suitable order, we can thus calculate both non-perturbative and perturbative contributions to the particle density, in particular for asymptotic times. \footnote{Although this procedure is presented as a perturbative expansion with respect to the oscillating field
%in the contribution to the total particle density,
we note that due to the source terms  in the differential equations, each level of the expansion can still include non-perturbative physics. And so are, for instance, quantum interference contributions an inherent aspect of the perturbative expansion at all levels. It is therefore important to interpret the results with caution, as only the total probability, summing over all contributions in $f$, represents a physical quantity.}

This procedure relies on numerically solving a comparatively simple system of differential equations, rather than evaluating more complex analytical structures. We will show below that it can provide greater physical insight on contributions to the particle distribution function from a fixed number of interactions with the secondary field. Moreover, the hierarchical quantum kinetic theory (HQKT) developed here is sufficiently general to function in a wide variety of different regimes of pair production, thus complementing \cite{PhysRevD.97.096004} and connecting with \cite{PhysRevD.80.111301, PhysRevD.100.116018}, while retaining the flexibility to go beyond these works. 
 
\section{Results and Discussion}
\label{sec:results}

\begin{figure}[t]
 \centering
  \begin{tikzpicture}[x=100,y=100]
    \node[anchor=north west] at (0,0) {
      \includegraphics[trim={0cm 0cm 3.5cm 0cm},clip,width=0.99\columnwidth]{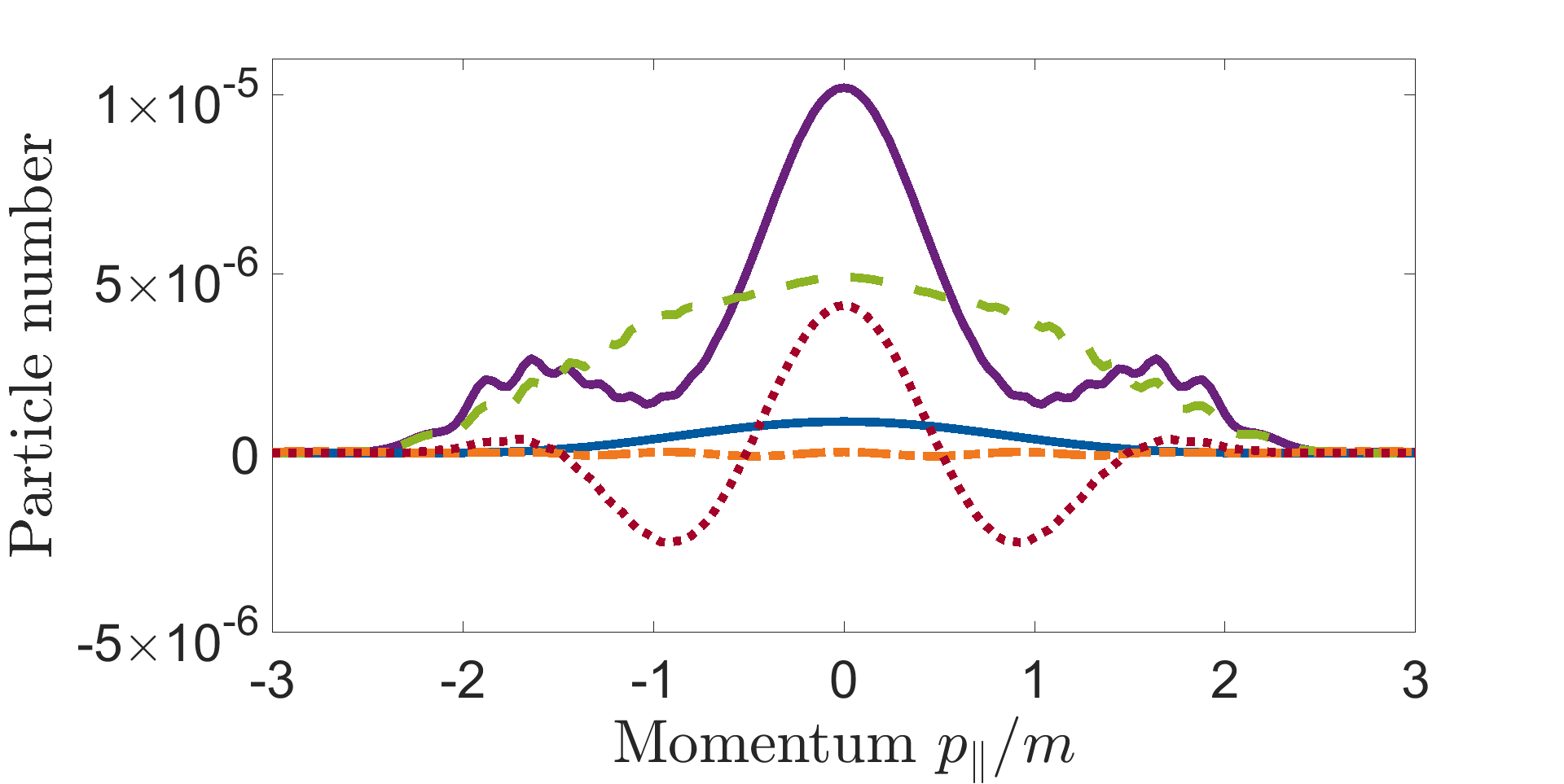}
    };
    \node[anchor=north west] at (2.02,-0.1) {
      \includegraphics[trim={0.0cm 0cm 0cm 0cm},clip,width=0.14\columnwidth]{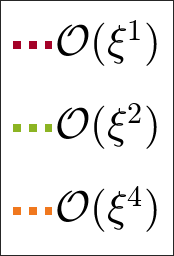}
    };    
    \node[anchor=north west] at (1.72,-0.1) {
      \includegraphics[trim={0.0cm 0cm 0.5cm 0cm},clip,width=0.125\columnwidth]{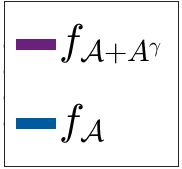}
    };    
    \end{tikzpicture}
  \caption{Spectral plot of the asymptotic particle number against the momentum $p_\parallel$,  with both vector potentials (purple) compared to the strong-field (blue) contribution,  with $p_\perp=0$. The dominant contributions to the full spectrum  come from absorbing one energy quantum (green) and its interference with the strong-field (red) (note this lowers the particle number for small $p_{\parallel}$). The orange curve corresponds to the perturbative contribution of order $\mathcal O(\xi^4)$, i.e. absorbing two quanta of energy. Field parameters: $\mathcal{E}=0.3,\ \varepsilon=0.05,\ \Omega=m$,\ $\mathcal{T} = \tau = 20/m$.}
  \label{fig1}
\end{figure} 

In past literature on the assisted Schwinger mechanism the primary focus has been on optimizing the secondary pulse profile to maximize the total particle yield, looking towards future experimental detection. In contrast, working at the spectral level permits  differentiation of pair production rates as a function of particle momenta and can identify effects that may have already been integrated out in the total yield.
%Secondly, new effects within assisted particle creation can be identified that may have already been integrated out at the level of the total yield.
As an example, one can consider the alternative perspective of identifying momenta at which \textit{no} pairs are created \cite{PhysRevD.108.036019}.

To illustrate the advantages of the proposed formalism we consider a strong electric pulse of field strength $0.3E_{\rm cr.}$, cf. $\mathcal{A}(t)$ in Eq.~\eqref{eq:A1}, superposed by a fast oscillating, but weak linearly polarized light field, cf. $A^{\gamma}(t)$ in Eq.~\eqref{eq:A2} with a field frequency of $\Omega=m$ and field strength $0.05 E_{\rm cr.}$. Due to its high field strength, the first field induces non-perturbative particle creation. The second field with a Keldysh parameter of $\gamma = 20$ is demonstrably within the perturbative regime. 

Calculation of the particle spectra for the combined field $\mathcal{A}(t)+A^{\gamma}(t)$, cf. Fig.~\ref{fig1}, reveals the non-linear nature of strong-field pair production. As expected, the number of created particle pairs $f_{\mathcal{A}+A^{\gamma}}$ is greater than the sum of the individual contributions $f_{\mathcal{A}} + f_{A^\gamma}$. However, the formalism developed here permits signals in the combined spectrum to be associated with the two fields separately, particularly multiple interactions with the oscillating field. We argue this can enhance understanding of the mechanisms governing the overall particle production. For example, Fig.~\ref{fig1} shows that for this scenario the most significant contributions are the first,  $\mathcal{O}(\xi^1)$, and second order terms, $\mathcal{O}(\xi^2)$, from the perturbative expansion. 
%\J{Very good}

To give a more accessible interpretation of the HQKT results, a semi-classical picture may be illustrative. From this perspective, a high-intensity background field enables the particle pairs to come into existence via quantum tunnelling through the rest-mass threshold energy barrier, while perturbative interactions with the oscillating field can cause the particles to gain energy and cross over the mass threshold of $2m$. When combined, the net effect is a conversion of energy from the electromagnetic field into fermion rest-mass and kinetic energy, with the assisting field lowering the effective barrier for tunnelling to occur.

In the specific case of Fig.~\ref{fig1}, we interpret the contribution $\mathcal{O}(\xi^2)$ to the result obtained as the electron-positron pair being "activated" through absorbing one quantum of energy from the oscillating field so that its effective tunneling mass is reduced by $\Omega$. Consequently, the effective threshold is only $m$, making subsequent quantum tunneling much easier.
% 
%\key{I would delete this paragraph (its confusing and we have to save words)}
%\J{\textbf{Since half of its rest mass has already been provided by the oscillating field} -- sorry, probably being slow -- how does a field strength of $0.05E_{cr}$ and frequency $\Omega = m$ lead to energy transfer of exactly $m/2$? First approximation: spatio-temporal constant field of that field strength over 1 Compton wavelength transfers $0.05m$.} \ck{Good question: Threshold is $2m$. Transfer of energy $\Omega$ leads to effective threshold of $2m-\Omega=m$. And that is half of where we started. In this picture field strength is irrelevant - see also photoelectric effect.} \\
%\ck{\sout{its effective tunneling mass is drastically reduced, making subsequent quantum tunneling much easier. As a result we find a strong contribution to the particle number at order $\xi^2$.}}
%\na{$\Omega\sim m$ in our case and that is why we could go beyond what Greger et all did, since their $\Omega$ is much smaller, is that true? This fits with quite well the energy transfer discussion. }
%
Furthermore, as this effect relies on a strong, slowly varying field as a background for the energy transfer, and that field is still capable of producing particles through quantum tunneling, we also identify interference between non-perturbative Schwinger and dynamically assisted particle production. As such, quantum interference at $\mathcal{O}(\xi)$ contribute significantly towards the spectrum.
 
\begin{figure}
 \centering
  \begin{tikzpicture}[x=100,y=100]
    \node[anchor=north west] at (0,0) {
      \includegraphics[trim={0cm 0cm 3.5cm 0cm},clip,width=0.99\columnwidth]{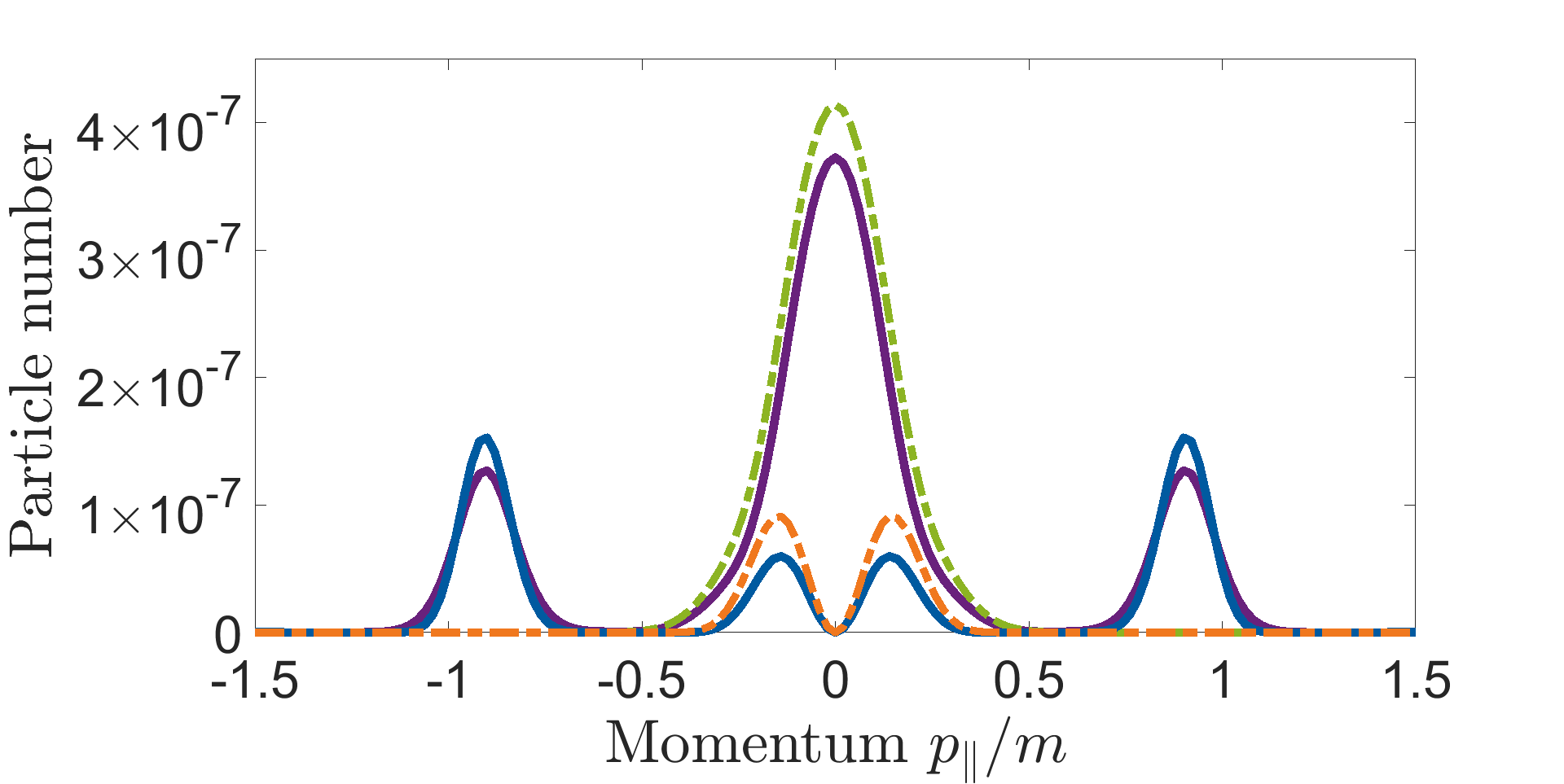}
    };
    \node[anchor=north west] at (1.68,-0.1) {
      \includegraphics[width=0.275\columnwidth]{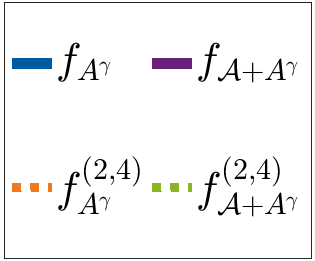}
    };
    \end{tikzpicture}
  \caption{Asymptotic particle number as a function of momentum $p_\parallel$ with $p_\perp=0$ for an oscillating field, $A^{\gamma}(t)$, only (blue and orange lines), and in conjunction with a weak, slowly varying field, $\mathcal{A}(t)+A^{\gamma}(t)$, (purple and green lines). In both cases, the total particle yield is shown as a solid line and the particle number at order $\mathcal O(\xi^4)$ are displayed as dashed lines. Field parameters: $\mathcal{E}=0.01,\ \varepsilon=0.1,\ \Omega=0.9m$,\ $\mathcal{T} = \tau = 20/m$. 
  %\key{Dangerous because different mechanism:} \J{Note the existence of momenta for which no pairs are created [Chris - check] (c.f. \cite{Ahmadiniaz:2022lea}).}
  }
  \label{fig2}
\end{figure}

The application of HQKT offers further insight when higher-order perturbations are significant. In this case, even a relatively weak, slowly varying field can significantly enhance pair production if the oscillating field "activates" the pair to be just below the threshold energy \cite{PhysRevD.100.116018}. In Fig.~\ref{fig2} the frequency of the oscillating field, $A^{\gamma}(t)$, is $\Omega = 0.9m$. Therefore, only the signal resulting from the absorption of three field quanta, associated with the peaks at $p_\parallel \sim \pm m$, is fully developed; pair production through the absorption of two field quanta is suppressed at $p_\parallel \sim \pm 0.1 m$. However, when an additional pulse, $\mathcal{A}(t)$, is applied the resulting particle spectrum exhibits a significant alteration, particularly in the vicinity of vanishing momentum.
 
Intuitively, the absorption of two field quanta results in an energy gain of $1.8m$, which in turn leads to an effective mass that is only ten percent of the particle pair mass in the context of quantum tunneling. Consequently, a much lower field strength, exemplified in this scenario by a pulse $E=0.01 E_{\rm cr.}$, can induce tunneling. In contrast, for the specific scenario shown in Fig.~\ref{fig2}, the particle distribution at large momenta $p_\parallel \approx \pm m$ is hardly affected by the addition of the auxiliary field, since the absorption of, e.g., three field quanta is already pushing the pair above the threshold.
  
Another crucial detail is that for purely oscillating fields, symmetry constraints in $n$-photon pair production can prevent the final particle distribution from being non-zero at certain momenta and for certain field frequencies. By simply adding a secondary field, this underlying symmetry is broken and, as a consequence, such restrictions are lifted. As a result, in Fig.~\ref{fig2} the particle spectra at $p_\parallel=0$ are qualitatively different.

\begin{figure}[t]
 \centering
  \begin{tikzpicture}[x=100,y=100]
    \node[anchor=north west] at (0,0) {
      \includegraphics[trim={0cm 0cm 3.5cm 0cm},clip,width=0.99\columnwidth]{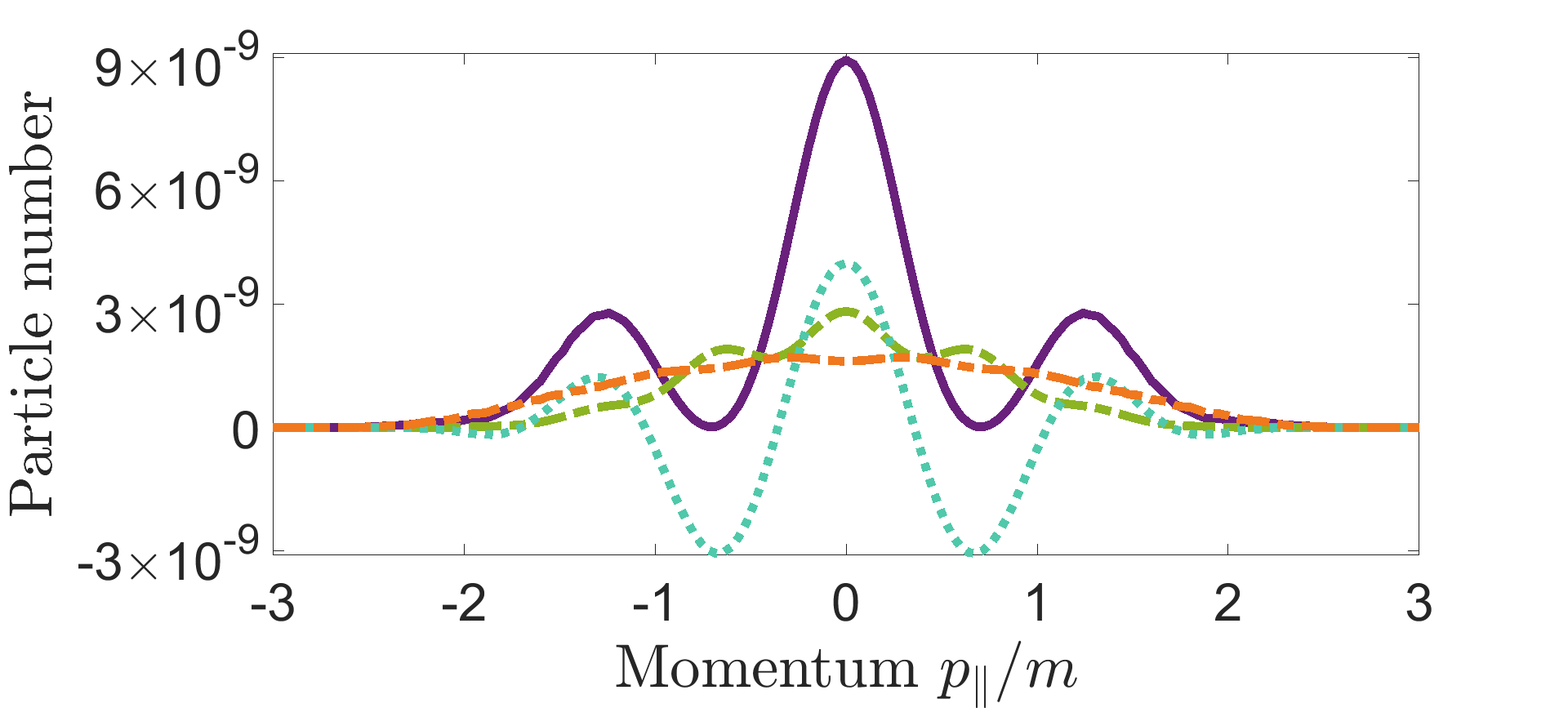}
    };
    \node[anchor=north west] at (1.68,-0.1) {
      \includegraphics[width=0.275\columnwidth]{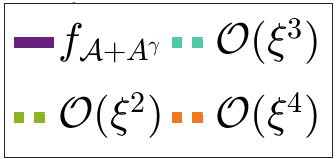}
    };
    \end{tikzpicture}
 \caption{Asymptotic particle number as a function of momentum $p_\parallel$ with $p_\perp=0$ for both fields, $\mathcal{A}(t) + A^{\gamma}(t)$. The total particle yield is shown as the solid purple line. In comparison, the contributions from orders $\mathcal O(\xi^2)$ and $\mathcal O(\xi^4)$ are displayed as green, dashed and orange, dashed lines, respectively. The most important strong-field interference term is of order $\xi^3$ and displayed in teal. Field parameters: $\mathcal{E}=0.15$, $\varepsilon=0.02,\ \Omega=0.7m$,\ $\mathcal{T} = \tau = 40/m$.}
 \label{fig3}
\end{figure}  

Access to the particle spectrum and the ability to distinguish contributions permit a more detailed examination of the role played by higher-order contributions in Schwinger pair production. So far, we have analyzed effects where tunneling particle creation was aided by an $n$-th order perturbative process, or vice versa. In contrast, in Fig.~\ref{fig3}, we have displayed results where it is the interplay of one- and two-energy packet absorption that drives particle creation. %Here the field frequency of the weak field is $\Omega=0.7m$, thus unassisted perturbative contributions would only be observable at the sixth order \J{Last 2 sentences not clearly related}.

In this case, with the frequency of the secondary field of $\Omega=0.7m$, the field $\mathcal{A}(t)$ occupies a "Goldilocks zone", exhibiting sufficient strength to provoke a sizeable increase in total particle yield compared to the unassisted scenario, while remaining too weak to entirely dominate the pair production process. Accordingly, assistance through the absorption of a single quantum of energy from the oscillating field yields a substantial contribution. However, another fraction originates from pairs that absorb two quanta of energy. Without the field $\mathcal{A}(t)$ the latter would be a less probable route to pair production due to its scaling by fourth order in the field strength. Nevertheless, the effective mass of such a pair is $2m_{\rm eff} = 2m - 1.4m$ and therefore sufficiently small that in the second stage of the formation process it has a distinct advantage over the alternative. 
%\J{Gorgeous}

Furthermore, interference has a substantial impact on the particle spectra, e.g., at $p_\parallel \approx \pm 0.72m$ in Fig.~\ref{fig3} the particle distribution is reduced by one order of magnitude. The latter aspect represents another significant advantage of HQKT, as this feature is not discernible in the total yield. To illustrate, in this specific scenario, roughly $99$ percent of the total yield is concentrated within orders $\xi^2$ and $\xi^4$. Additional figures showing how the relative contributions from the competing pair creation mechanisms varies with field strengths are provided in the Supplemental (see Fig. 4.).

\vspace{-0.5cm}
\section{Conclusion}
\label{sec:conc}

We have shown how ideas from perturbation theory can be applied at the level of \textit{probabilities} within the mean-field approximation in relativistic quantum kinetic theory, without sacrificing access to non-perturbative physics. This led to a hierarchical approach to quantum kinetic theory characterised by an infinite tower of quantum kinetic equations. These can be solved order by order in a natural expansion related to a secondary (in our case oscillating) field. This approach gives physically intuitive results at each order in that field's "classical" nonlinearity parameter.

In this regard, we have analyzed the momentum spectra of created particles in terms of both perturbative and non-perturbative contributions. We have shown how this approach can be applied in three examples of dynamical assistance to the particle pair creation process. Not only can the overall particle yield be increased, but we find that the superposition of an oscillating field with a weak, slowly varying pulse can lead to particle production in an otherwise forbidden momentum region. Conversely, quantum interference between different paths of assistance -- and indeed in interplay with the strong background field -- has the potential to alter the spectrum, creating regions where the occupation probability is drastically modified. In the context of strong field QED, analogous interference effects that contribute to back reaction onto the background field were studied for the radiated waveform in \cite{Copinger:2024pai}.

The method of splitting the particle distribution function into components and then expanding the transport equations is also applicable to more complex scenarios and field geometries. In this respect, the approach introduced in this article may well become the starting point for further explorations regarding new perturbative approaches to theoretical calculations in strong-field quantum physics. 

\section*{Acknowledgments}
  
The authors thank Ralf Schützhold for interesting discussions. CK particularly appreciates exchanges with Friedemann Queisser and Stefan Evans. Numerical computations have been performed on the Hemera-Cluster in Dresden.

%%%%%%%%%%%%%%%%%%%%%%%%%%%%%%%%%%%%%%%%%%%%%%%%%%%%%%%%%%%%%%%%%%%%%%%%%%%%%%%%%

%\paragraph{Bibliography} 

%\newpage 
\interlinepenalty=10000

%%%%%%%%%%%%%%%%%%%%%%%%%%%%%%%%%%%%%%%%%%%%%%%%%%%%%%%%%%%%%%%%%%%%%%%%%%%%%%%%%

%%%%%%%%%% Merge with supplemental materials %%%%%%%%%%
\pagebreak

%%%%%%%%%% Merge with supplemental materials %%%%%%%%%%
%%%%%%%%%% Prefix a "S" to all equations, figures, tables and reset the counter %%%%%%%%%%
\setcounter{equation}{0}
\setcounter{figure}{0}
\setcounter{table}{0}
\setcounter{page}{1}
\makeatletter
\renewcommand{\theequation}{S\arabic{equation}}
\renewcommand{\thefigure}{S\arabic{figure}}
\renewcommand{\bibnumfmt}[1]{[S#1]}
\renewcommand{\citenumfont}[1]{S#1}
%%%%%%%%%% Prefix a "S" to all equations, figures, tables and reset the counter %%%%%%%%%%

 \begin{widetext}

    \begin{center}
    \textbf{\large Supplemental}
    \end{center}

    %\begin{center}
    \noindent
    The supplementary file provides a more comprehensive theoretical background, with a particular focus on the seemingly different formalism discussed in the article. It also presents technical aspects regarding the numerical part of the paper that may be of interest to readers with a technical background. These include further test cases, convergence plots, and technical data in tabular form. %A functioning code sample in Matlab is provided in "pairproduction\_v4.m".
    %\end{center} 

 \end{widetext}

\appendix

\section{Recap of the article}

For the sake of convenience, the most important quantities from the main text are shown here again. First, the critical field strength $E_{\rm cr.} \approx 1.3 \times 10^{18}$~V/m and the electron mass $m \approx 511$ keV. Additionally, we use units where $c=\hbar=1$.

The key element of the article was the decomposition of the gauge potential into a pulse field, treated non-perturbatively, and an assisting field, which we treat perturbatively: $\mathbf{A} = \boldsymbol{\mathcal{A}} + \mathbf{A}^{\gamma}$. Since in the main article we only discuss fields with linear polarization, we omit the vector notation in the following. The pulse which is taken as is exhibits a field strength $\mathcal{E}$ and a pulse duration $\mathcal{T}$, e.g.,
\begin{equation}
  \mathcal{A}(t) = \mathcal{E} E_{\rm cr.} \, t \, \exp(-t^2/\mathcal{T}^2). 
  \label{eq:A1} 
\end{equation}
An oscillating field, which is taken into account perturbatively, is characterized through field strength  $\varepsilon$, pulse envelope $\tau$, frequency $\Omega$, and profile
\begin{equation}
 A^{\gamma}(t) = \varepsilon\, E_{\rm cr.} /\Omega \,  \exp(-t^2/\tau^2) \ \sin(\Omega t). \label{eq:A2}
\end{equation}
The quantum nonlinearity parameter is $\xi = e \varepsilon E_{\rm cr.}/m\Omega$.

%The background pulse has field strength $\mathcal{E}$ and pulse duration $\mathcal{T}$,
%\begin{equation}
% A_1(t) = \mathcal{E} E_{\rm cr.} \, t \, \exp(-t^2/\mathcal{T}^2), 
% \label{eq:A1} \\
%\end{equation}
%and the  oscillating field has field strength $\varepsilon$, pulse envelope duration $\tau$ %and frequency $\Omega$,
%\begin{equation}
% A_2(t) = \varepsilon\, E_{\rm cr.} /\Omega \,  \exp(-t^2/\tau^2) \ \sin(\Omega t). 
% \label{eq:A2}
%\end{equation}
%The quantum nonlinearity parameter is $\xi = e \varepsilon E_{\rm cr.}/m\Omega$.

Finally, the basic differential equations for the Wigner components (that we shall derive here) are
\begin{alignat}{8}
    & \dot{\mathbbm{s}} && && -2 p_\parallel \mathbbm{v}_\perp && && +2 q_\perp \mathbbm{v}_\parallel && && &&= 0, \label{eq:w1a} \\  
    & \dot{\mathbbm{v}}_\parallel && && && && -2 q_\perp \mathbbm{s} && &&  &&= -2m\mathbbm{v}_\perp , \\   
    & \dot{\mathbbm{v}}_\perp && && +2 p_\parallel \mathbbm{s} && && && && &&= +2m\mathbbm{v}_\parallel, \label{eq:w1b}
  \end{alignat} 
for particle momenta $q_\perp$ and $p_\parallel = q_\parallel -eA$ depending on the alignment with the vector potential, with vacuum initial conditions 
\begin{equation}
 \mathbbm{s}^{(0)} = -m/\omega ,\ \mathbbm{v}_\parallel^{(0)} = -q_\parallel/\omega ,\ \mathbbm{v}_\perp^{(0)} = -q_\perp/\omega.  
 \label{eq:A6}
\end{equation}

\begin{figure}[t]
 \includegraphics[width=0.95\columnwidth]{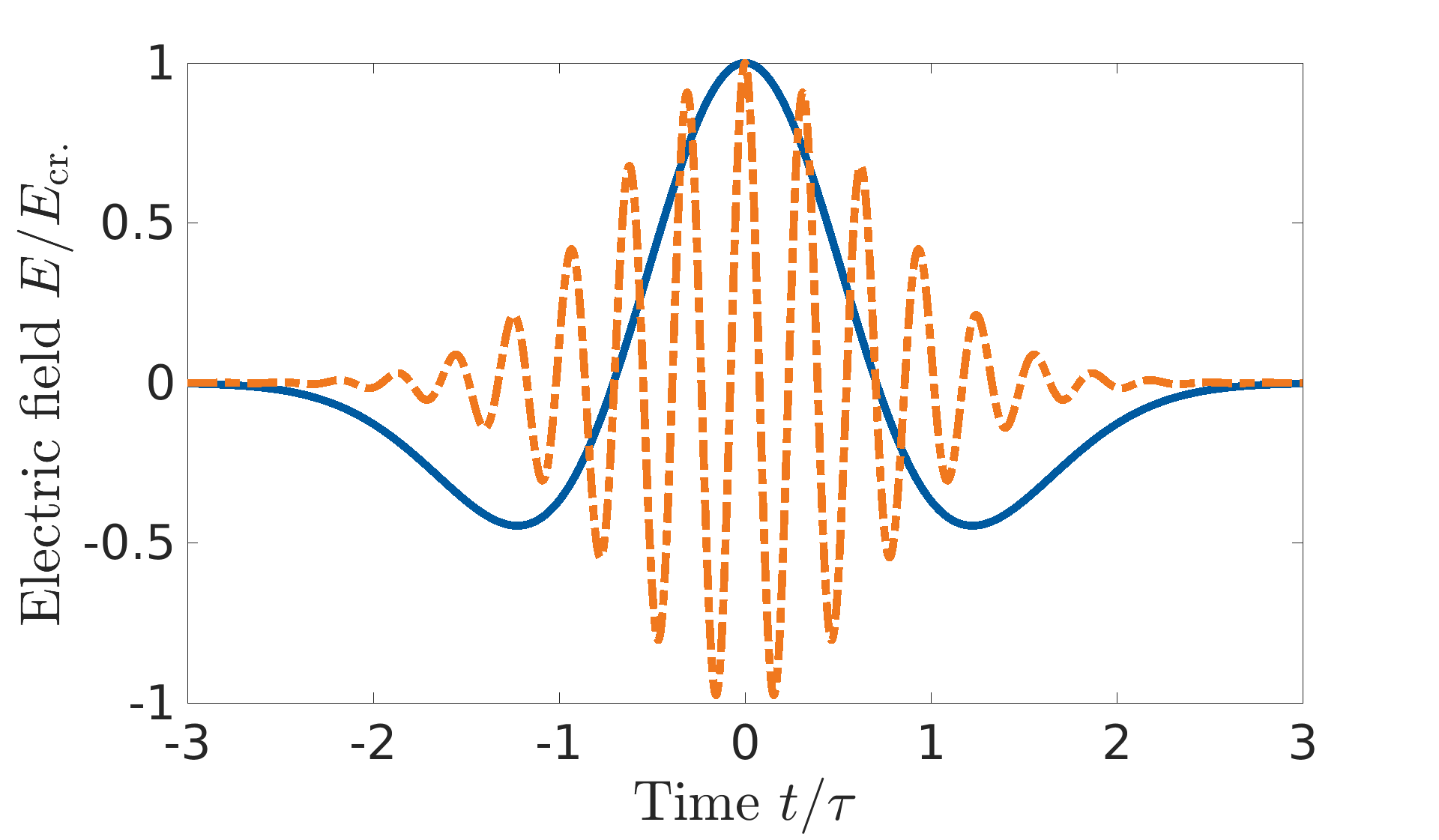}
 \caption{Electric fields as a function of time derived from the gauge potentials defined in the text, $\mathcal{A}(t)$ in blue and $A^\gamma(t)$ in orange. For demonstration purposes we have set the pulse parameters to $\mathcal{E}=\varepsilon=-1,\ \omega=1.0m$,\ $\mathcal{T} = \tau = 20/m$.}
 \label{fig:A}
\end{figure}

%The particle distribution is defined in terms of mass, $\mathbbm{s}$, and current, $\mathbbm{v}_\parallel$ and $\mathbbm{v}_\perp$, distributions
%\begin{equation}
% \omega f(t, \mathbf{p}) = m \mathbbm{s} + p_\parallel \mathbbm{v}_\parallel + p_\perp \mathbbm{v}_\perp.
% \label{eq:f}
%\end{equation} 

\newpage
\section{Quantum electrodynamics in \\ the phase-space formalism}

This section provides a more detailed account of the theoretical formalism employed in the article. We begin by defining the equal-time Wigner function within mean-field theory as the relativistic complement to ordinary kinetic theories. It is defined in phase-space, with an explicit time dependence, as the background electric field drives the time-dependence of quantized Dirac spinors. As a result, in the limit of $\hbar \to 0$, the Vlasov as well as the Bargmann-Michel-Telegdi equations are recovered. We use vacuum initial conditions, thereby enabling interpretation of the results in terms of the vacuum state being driven out of equilibrium by the background electric field.

It should be noted that in the main article, the derivation was presented in reverse order. Initially, the equations for the distribution function were stated, and subsequently, a method for obtaining a set of differential equations was demonstrated. This approach was employed because it is anticipated that readers less familiar with the Wigner formalism in second quantization may be more familiar with the dynamics of open quantum systems or quantum kinetic theories.

For readers interested in a more fundamental account of the formalism, we direct them to the reviews and reports, \cite{Vasak, PhysRevD.105.076017, Zhuang, Carruthers, PhysRevD.44.1825, Hidaka:2022dmn, PhysRevD.57.6525, Ochs}. The following derivation is mostly based on \cite{PhysRevD.44.1825}.

The equal-time Wigner operator is defined as follows: 
\begin{multline}
 \hat{\mathbbm{w}}_{\alpha \beta} \left( t, \boldsymbol{x}, \boldsymbol{p} \right) = \\
 \frac{1}{2} \int {\rm d}^3 s \ e^{- \mathrm{i} \boldsymbol{p} \cdot \boldsymbol{s}} \ e^{-\mathrm{i}e \int_{-1/2}^{1/2} {\rm d} \xi \ \boldsymbol{A} \left(t, \boldsymbol{x} + \xi \boldsymbol{s} \right) \cdot \boldsymbol{s} } \\
 \times \left[ \hat{\bar {\psi}}_\beta  \left( t, \boldsymbol{x} - \boldsymbol{s}/2 \right), \hat{\psi}_\alpha \left( t, \boldsymbol{x} + \boldsymbol{s}/2 \right) \right], 
\end{multline}
with quantized spinor fields $\hat{\bar {\psi}}$ and $\hat{{\psi}}$ and a classical, in principle spatially inhomgoeneous, gauge potential $\boldsymbol{A}$ that appears integrated along a straight line between the endpoints. The Wigner operator is defined with respect to a spinor basis (indices $\alpha$ and $\beta$), and is thus matrix-valued. It is worth pointing out that a mean-field approximation of Hartree type has already been employed to arrive at this point, wherein quantum fluctuations of the electric field have been disregarded, so as to leave the gauge field a c-number representing a fixed, classical background.

In order to eventually obtain physically accessible observables, the average of the Wigner operator is taken. With regard to pair creation from the quantum vacuum, the expectation values are calculated with respect to the vacuum state
\begin{equation}
 \mathbbm{w} \left( t, \mathbf{x} , \mathbf{p} \right) =
 {\scriptsize  
  \begin{matrix}
   \\
   {\rm in}
  \end{matrix}
 }
 \langle \Omega \lvert \hat{\mathbbm{w}}  \left( t, \mathbf{x} , \mathbf{p} \right) \rvert \Omega \rangle  
 {\scriptsize  
  \begin{matrix}
   \\
   {\rm in}
  \end{matrix}
 }\,. 
\end{equation}
Particular emphasis is placed on the fact that the states are taken as initial states, i.e. before coupling to the gauge potential. 

To proceed, the matrix-valued Wigner function $\mathbbm{w} \left( t, \mathbf{x} , \mathbf{p} \right)$ is expanded into a complete set of matrices (the gamma matrices in $(3+1)$ dimensions), e.g., mass density $\mathbbm{s}$, which shifts the focus to Wigner components,
\begin{equation}
{\mathbbm{w}} = \frac{1}{4} \left(\mathbbm{1} {\mathbbm{s}} +  \gamma^{\mu}  {\mathbbm{v}}_{\mu}  + \sigma^{\mu \nu} {\mathbbm{t}}_{\mu \nu} + \gamma^{\mu} \gamma_5 { \mathbbm{a}}_{\mu} + {\rm i} \gamma_5 {\mathbbm{p}} \right). 
\end{equation}
The Wigner function satisfies its own Dirac equation, with (non-local) covariant momentum and derivative operators that are dressed by the background field, see \cite{Vasak}. At the level of the Wigner components, this results in the following set of quantum transport equations
 \begin{alignat}{4}
    & D_t \mathbbm{s}     && && -2 \boldsymbol{\Pi} \cdot \boldsymbol{\mathbbm{t_1}} &&= 0, \label{eq:pair_D3} \\
    & D_t \mathbbm{p} && && +2 \boldsymbol{\Pi} \cdot \boldsymbol{\mathbbm{t_2}} &&= -2m\mathbbm{a}_\mathbb{0},  \\
    & D_t \mathbbm{v}_\mathbb{0} &&+ \boldsymbol{D} \cdot \boldsymbol{\mathbbm{v}} && &&= 0, \\
    & D_t \mathbbm{a}_\mathbb{0} &&+ \boldsymbol{D} \cdot \boldsymbol{\mathbbm{a}} && &&=+ 2m\mathbbm{p}, \\    
    & D_t \boldsymbol{\mathbbm{v}} &&+ \boldsymbol{D} \ \mathbbm{v}_\mathbb{0} && +2 \boldsymbol{\Pi} \times \boldsymbol{\mathbbm{a}} &&= -2m\boldsymbol{\mathbbm{t_1}}, \\    
    & D_t \boldsymbol{\mathbbm{a}} &&+ \boldsymbol{D} \ \mathbbm{a}_\mathbb{0} && +2 \boldsymbol{\Pi} \times \boldsymbol{\mathbbm{v}} &&= 0, \\
    & D_t \boldsymbol{\mathbbm{t_1}} &&+ \boldsymbol{D} \times \boldsymbol{\mathbbm{t_2}} && +2 \boldsymbol{\Pi} \ \mathbbm{s} &&= +2m\boldsymbol{\mathbbm{v}}, \\    
    & D_t \boldsymbol{\mathbbm{t_2}} &&- \boldsymbol{D} \times \boldsymbol{\mathbbm{t_1}} && -2 \boldsymbol{\Pi} \ \mathbbm{p} &&= 0,  \label{eq:pair_D4}
  \end{alignat} 
with integro-differential operators
  \begin{alignat}{6}
     & D_t && = \partial_t &&+ e &&\int_{-1/2}^{1/2} {\rm d} \xi &&\boldsymbol{E} \left( \boldsymbol{x}+{\rm i} \xi \boldsymbol{\nabla}_p,t \right) && ~ \cdot \boldsymbol{\nabla}_p, \label{eq:pair_D1} \\[-2mm]
     & \boldsymbol{D} && = \boldsymbol{\nabla}_x &&+ e &&\int_{-1/2}^{1/2} {\rm d} \xi &&\boldsymbol{B} \left( \boldsymbol{x}+{\rm i} \xi \boldsymbol{\nabla}_p,t \right) &&\times \boldsymbol{\nabla}_p, \\[-2mm]
     & \boldsymbol{\Pi} && = \boldsymbol{p} &&- {\rm i} e &&\int_{-1/2}^{1/2} {\rm d} \xi \,\xi &&\boldsymbol{B} \left( \boldsymbol{x}+{\rm i} \xi \boldsymbol{\nabla}_p,t \right) &&\times \boldsymbol{\nabla}_p. \label{eq:pair_D2} 
  \end{alignat}    
 
The energy in the fermionic sector is given by 
\begin{multline}
 \mathfrak E = \int {\rm d}^3 x \ \langle \mathfrak T^{00} (t,\mathbf{x}) \rangle \\
 = \int \dfrac{{\rm d}^3 x \ {\rm d}^3 p}{(2 \pi)^3} \ m \mathbbm{s} (t,\mathbf{x}, \mathbf{p}) + \mathbf{p} \cdot \boldsymbol{\mathbbm{v}} (t, \mathbf{x}, \mathbf{p}),
\end{multline}  
where $\mathfrak T$ is the energy-momentum tensor.
Correspondingly, the particle density is given by normalizing the total energy by the one-particle energy $\omega = \sqrt{m^2 + \mathbf{p}^2}$. In the absence of any field, equilibrium vacuum initial conditions can be expressed as follows (c.f. (\ref{eq:A6})):
\begin{equation}
 \mathbbm{s}_{\rm vac} = -\dfrac{m}{2\omega}, \quad \boldsymbol{\mathbbm{v}}_{\rm vac} = -\dfrac{\mathbf{p}}{2\omega}.
\end{equation}

In the case of a purely homogeneous background and homogeneous initial and boundary conditions, the quantum transport equations are significantly simplified. In particular, for a purely time-dependent electric field, the integro-differential operators become the familiar kinetic momentum and spatial derivative operators. For the sake of simplicity, we focus on linearly polarized fields, for which only one vector component of the electric field is non-zero, e.g., $\mathbf{E}(t) = (E_x(t),0,0)$. Through minimal coupling 
\begin{align}
 & p_x = q_x - e A_x(t), \, && p_y = q_y, \, && p_z=q_z,
\end{align}
we then obtain the following set of differential equations for an Abelian plasma with vacuum initial conditions
\begin{alignat}{8}
    & \partial_t \mathbbm{s} && && -2 q_x \mathbbm{v}_3 && && +2 q_z \mathbbm{v}_1 && + 2e A_x \mathbbm{v}_3 && &&= 0, \\  
    & \partial_t \mathbbm{v}_1 && && && && -2 q_z \mathbbm{s} && &&  &&= -2m\mathbbm{v}_3 ,  \\   
    & \partial_t \mathbbm{v}_3 && && +2 q_x \mathbbm{s} && && && -2e A_x \mathbbm{s} && &&= +2m\mathbbm{v}_1.
  \end{alignat}  
All other Wigner components vanish entirely. At this stage, it is possible to modify the notation in order to reflect the fact that the electric field (and, by extension, the vector potential) points in a single direction, and that all vector quantities are either parallel or perpendicular to it \eqref{eq:w1a}-\eqref{eq:w1b}. 

By combining the Wigner components $\mathbbm{s}$, $\mathbbm{v}_1$ and $\mathbbm{v}_3$ (or more appropriately $\mathbbm{s}$, $\mathbbm{v}_\parallel$ and $\mathbbm{v}_\perp$) in a specific order, 
%\begin{alignat}{8}
% \mathbbm{s} = & \frac{m}{\omega} \, f && -\frac{m \, (q_\parallel - eA)}{\varepsilon_\perp \omega} \, \mathcal{G} && - \frac{q_\perp}{\varepsilon_\perp} \, \mathcal{H}, \\
% \mathbbm{v}_\parallel = & \frac{q_\parallel - eA}{\omega} \, f && + \frac{\varepsilon_\perp}{\omega} \, \mathcal{G} &&, \\ 
% \mathbbm{v}_\perp = & \frac{q_\perp}{\omega} \, f && -\frac{q_\perp \, (q_\parallel - eA)}{\varepsilon_\perp \omega} \, \mathcal{G} && + \frac{m}{\varepsilon_\perp} \, \mathcal{H},   
%\end{alignat}  
\begin{alignat}{8}
 & \mathbbm{s} & = && \frac{m}{\omega} \, f && -\frac{m \, (q_\parallel - eA)}{\varepsilon_\perp \omega} \, \mathcal{G} \phantom{,} && - \frac{q_\perp}{\varepsilon_\perp} \, \mathcal{H}, \\
 & \mathbbm{v}_\parallel & = && \frac{q_\parallel - eA}{\omega} \, f && + \frac{\varepsilon_\perp}{\omega} \, \mathcal{G}, && \\ 
 & \mathbbm{v}_\perp & = && \frac{q_\perp}{\omega} \, f && -\frac{q_\perp \, (q_\parallel - eA)}{\varepsilon_\perp \omega} \, \mathcal{G} \phantom{,} && + \frac{m}{\varepsilon_\perp} \, \mathcal{H},   
\end{alignat}  
we obtain the set of equations
\begin{alignat}{8}
    & \partial_t f && && = Q \, \mathcal{G}, && && && && && \\  
    & \partial_t {\mathcal {G}} && && = Q \, \left( 1- f \right) && - 2 \omega \mathcal{H} && && && &&, \\   
    & \partial_t {\mathcal {H}} && && = && + 2 \omega \, \mathcal{G} && && && &&, 
  \end{alignat}
already incorporating the initial conditions into the differential equation such that initially $f,\ \mathcal{G}$ as well as $\mathcal{H}$ are all zero. The auxiliary quantities read 
\begin{align}
 Q &= \frac{e E(t) \, \varepsilon_\perp}{\omega^2}, \label{eq:Qew1} \\ 
 \varepsilon_\perp &= \sqrt{m^2 + q_\perp^2},\\ 
 \omega^2 &= m^2 + q_\perp^2 + \left( q_\parallel - eA(t) \right)^2.  
 \label{eq:Qew3}
\end{align} 
For the sake of completeness, the new quantities $f,\ \mathcal{G}$ and $\mathcal{H}$ are given in terms of the Wigner components
%\begin{alignat}{8}
% f = & \frac{m}{\omega} \, \mathbbm{s} && +\frac{q_\parallel-eA}{\omega} \, \mathbbm{v}_\parallel && +\frac{q_\perp}{\omega} \, \mathbbm{v}_\perp, \\
% \mathcal{G} = & -\frac{m \, (q_\parallel - eA)}{\varepsilon_\perp \omega} \, \mathbbm{s} && + \frac{\varepsilon_\perp}{\omega} \, \mathbbm{v}_\parallel && -\frac{q_\perp \, (q_\parallel - eA)}{\varepsilon_\perp \omega} \mathbbm{v}_\perp, \\ 
% \mathcal{H} = & -\frac{q_\perp}{\varepsilon_\perp} \, s && && + \frac{m}{\varepsilon_\perp} \, \mathbbm{v}_\perp.   
%\end{alignat}  
\begin{alignat}{8}
 & f = & \frac{m}{\omega} \, \mathbbm{s} && \, + \, && \frac{q_\parallel-eA}{\omega} \, \mathbbm{v}_\parallel && +\frac{q_\perp}{\omega} \, \mathbbm{v}_\perp, \\
 & \mathcal{G} = & -\frac{m \, (q_\parallel - eA)}{\varepsilon_\perp \omega} \, \mathbbm{s} && \, + \, && \frac{\varepsilon_\perp}{\omega} \, \mathbbm{v}_\parallel && -\frac{q_\perp \, (q_\parallel - eA)}{\varepsilon_\perp \omega} \mathbbm{v}_\perp, \\
 & \mathcal{H} = & -\frac{q_\perp}{\varepsilon_\perp} \, \mathbbm{s} && && && + \frac{m}{\varepsilon_\perp} \, \mathbbm{v}_\perp.   
\end{alignat}  

\section{Quantum electrodynamics and \\ Bogoliubov formalism}

As an alternative to the derivation of a relativistic quantum transport theory presented in the last section, one may use Bogoliubov theory in the context of relativistic quantum electrodynamics in order to describe pair production. The idea is that for a vacuum subjected to a time-dependent "quench" by an electric field, single fermions are not representative of the excitations of the quantum system. Instead, pairs of fermions, e.g., an electron-positron, are the elementary objects. Bogoliubov theory deals with this discrepancy by translating the particle operators (before the electric field is applied) into new quasi-particle operators that already take into account the pair excitations. As such, for some this approach might provide a more intuitive progression from the Dirac equation to the quantum Boltzmann equation, and hence to the time dynamics of the electron-positron particle distribution.

To begin with, the second order Dirac equation expanded into a suitable spinor basis is given by
\begin{equation}
 \left( \partial_t^2 + \omega^2 + {\rm i}e E(t) \right) \chi(t) = 0,
\end{equation} 
with mode functions $\chi(t)$, electric field $E(t) = - \dot {A}(t)$ and where again $\omega^2 = m^2 + q_\perp^2 + \left( q_\parallel - eA(t) \right)^2$.   

Being a second order differential equation there are two solutions to the quadratic Dirac equation, which in turn can be used to expand the corresponding Dirac spinor in terms of fermionic, time-independent creation and annihilation operators ($\hat a,\ \hat a^\dag,\ \hat b,\ \hat b^\dag$). Hence the term mode functions. In a vacuum, i.e. before an electric field is applied, the mode functions represent plane waves and therefore these operators correspond to the states of free electrons and positrons respectively. However, in the presence of an electric field the solutions to the Dirac equation are not plane waves. A description in terms of such single-particle operators would therefore lead to a Hamiltonian with off-diagonal elements and a non-trivial particle interpretation.

To remedy this problem and therefore to diagonalize the Hamiltonian a single-particle interpretation is abandoned in favour of a quasi-particle representation. As such, new time-dependent creation and annihilation operators are introduced ($\hat A(t),\ \hat A(t)^\dag,\ \hat B(t),\ \hat B(t)^\dag$), which are related to the initial operators by a Bogoliubov transformation
\begin{align}
\hat A(t) = \alpha(t) \hat  a - \beta(t)^\ast \hat b^\dag,\\    
\hat B(t)^\dag = \beta(t) \hat a + \alpha(t)^\ast \hat b^\dag.
\end{align}
The complex coefficients $\alpha$ and $\beta$ are time-dependent and fulfill the condition $\lvert \alpha(t) \rvert^2 + \lvert \beta(t) \rvert^2 = 1$. Correspondingly, Dirac spinors are expanded in this quasi-particle basis and in terms of the time-dependent quasi-particle ladder operators. As such, the (adiabatic) mode functions in the quasi-particle basis are of the form
\begin{equation}
 \phi(t)^{(\pm)} \propto e^{\mp {\rm i} \Theta(t,t',\mathbf{q})},\ \text{with} \, \Theta(t,t',\mathbf{q}) = \int_t^{t'} {\rm d}t'' \, \omega(t'',\mathbf{q}),  
\end{equation}
where we have reintroduced labels for the momenta $\mathbf{q}$.
Hence, knowledge of the mode functions $\chi(t)$ and $\phi(t)$ translates into coupled differential equations for the Bogoliubov coefficients 
\begin{align}
 \partial_t \alpha(t, \mathbf{q}) & = + \frac{1}{2} Q(t, \mathbf{q}) \, \beta(t, \mathbf{q}) \, e^{+2 {\rm i} \Theta(t,t',\mathbf{q})},\\
 \partial_t \beta(t, \mathbf{q}) & = -\frac{1}{2} Q(t, \mathbf{q}) \, \alpha(t, \mathbf{q}) \, e^{-2 {\rm i} \Theta(t,t',\mathbf{q})},
\end{align}
where again $Q, \, \varepsilon_\perp$ and $\omega^2$ are defined as in Eqs.~\eqref{eq:Qew1}-\eqref{eq:Qew3}.
%\begin{align}
% Q &= \frac{e E(t) \, \varepsilon_\perp}{\omega^2},\\ 
% \varepsilon_\perp &= \sqrt{m^2 + q_\perp^2},\\ 
% \omega^2 &= m^2 + q_\perp^2 + \left( q_\parallel - eA(t) \right)^2.    
%\end{align}

The particle pair distribution function (with proper normalization) is given by
\begin{equation}
 f \left( t , \mathbf{q} \right) =
 {\scriptsize  
  \begin{matrix}
   \\
   {\rm in}
  \end{matrix}
 }
 \langle \Omega \lvert \hat A(t)^\dag \hat A(t) \rvert \Omega \rangle  
 {\scriptsize  
  \begin{matrix}
   \\
   {\rm in}
  \end{matrix}
 }
 = \lvert \beta \left( t , \mathbf{q} \right) \rvert^2, 
\end{equation}
assuming vacuum initial conditions 
\begin{equation}
{\scriptsize  
  \begin{matrix}
   \\
   {\rm in}
  \end{matrix}
 }
 \langle \Omega \lvert \hat a^\dag \hat a \rvert \Omega \rangle  
 {\scriptsize  
  \begin{matrix}
   \\
   {\rm in}
  \end{matrix}
 }=
 {\scriptsize  
  \begin{matrix}
   \\
   {\rm in}
  \end{matrix}
 }
 \langle \Omega \lvert \hat b^\dag \hat b \rvert \Omega \rangle  
 {\scriptsize  
  \begin{matrix}
   \\
   {\rm in}
  \end{matrix}
 }    
 = 0.
\end{equation}
Solving for $\beta$ therefore yields the integro-differential representation of the quantum Vlasov equation
\begin{alignat}{8}
\label{eqdtf}
 &\partial_t f(t, \mathbf{q}) = \\
 &Q(t, \mathbf{q}) \int_{t_{\rm i}}^t {\rm d}t'\, Q(t', \mathbf{q}) \left( 1 - f(t', \mathbf{q})\right) \cos \left(2 \Theta (t, t', \mathbf{q}) \right), \notag
\end{alignat}
with initial time $t_{\rm i}$ and phase $\Theta(t, t', \mathbf{q}) = \int_{t}^{t'} {\rm d}t''\, \omega(t'', \mathbf{q})$. 

In order to express this equation in differential form, it is necessary to introduce the quantities
\begin{alignat}{8}
 &G(t, \mathbf{q}) = \int_{t_{\rm i}}^t {\rm d}t'\, Q(t', \mathbf{q}) \left( 1 - f(t', \mathbf{q})\right) \cos \left(2 \Theta (t, t', \mathbf{q}) \right), \\
 &H(t, \mathbf{q}) = \int_{t_{\rm i}}^t {\rm d}t'\, Q(t', \mathbf{q}) \left( 1 - f(t', \mathbf{q})\right) \sin \left(2 \Theta (t, t', \mathbf{q}) \right).
\end{alignat}
In doing so we find that Eq.~(\ref{eqdtf}) is equivalent to
\begin{alignat}{8}
    & \partial_t f && && = Q \, G, && && && && && \\  
    & \partial_t G && && = Q \, \left( 1- f \right) && - 2 \omega H && && && &&, \\   
    & \partial_t H && && = && + 2 \omega G && && && &&,
  \end{alignat}
which, when identifying $G=\mathcal{G}$ and $H=\mathcal{H}$, precisely reproduces the differential equation previously derived through application of the Wigner formalism. 

Conversely, if we were to reformulate the entire system in terms of kinetic momenta, we would arrive at the quantum kinetic equations
\begin{alignat}{8}
    & ( \partial_t && + e E_x(t) \, \partial_{p_{\parallel}} ) f && = Q \, \mathcal{G}, && && && && && \\  
    & ( \partial_t && + e E_x(t) \, \partial_{p_{\parallel}} ) \mathcal{G} && = Q \, \left( 1- f \right) && - 2 \omega \mathcal{H} && && && &&, \\   
    & ( \partial_t && + e E_x(t) \, \partial_{p_{\parallel}} ) \mathcal{H} && = && + 2 \omega \, \mathcal{G} && && && &&.
  \end{alignat}
In conclusion, there are multiple approaches to obtaining the quantum Vlasov equation for linearly polarized, purely time-dependent electric fields. However, to enable the perturbative expansion as demonstrated in the article, it is necessary that the expansion parameter be linear in the differential equations. Neither the integral form of the Vlasov equation nor the frequently utilized set of equations in terms of the distribution function $f(t, \mathbf{q})$ can fulfill this requirement. This underscores the importance of working with the set \eqref{eq:w1a}-\eqref{eq:w1b}.

\section{Additional Figures and Data}

\begin{figure}[b]
 \includegraphics[width=0.95\columnwidth]{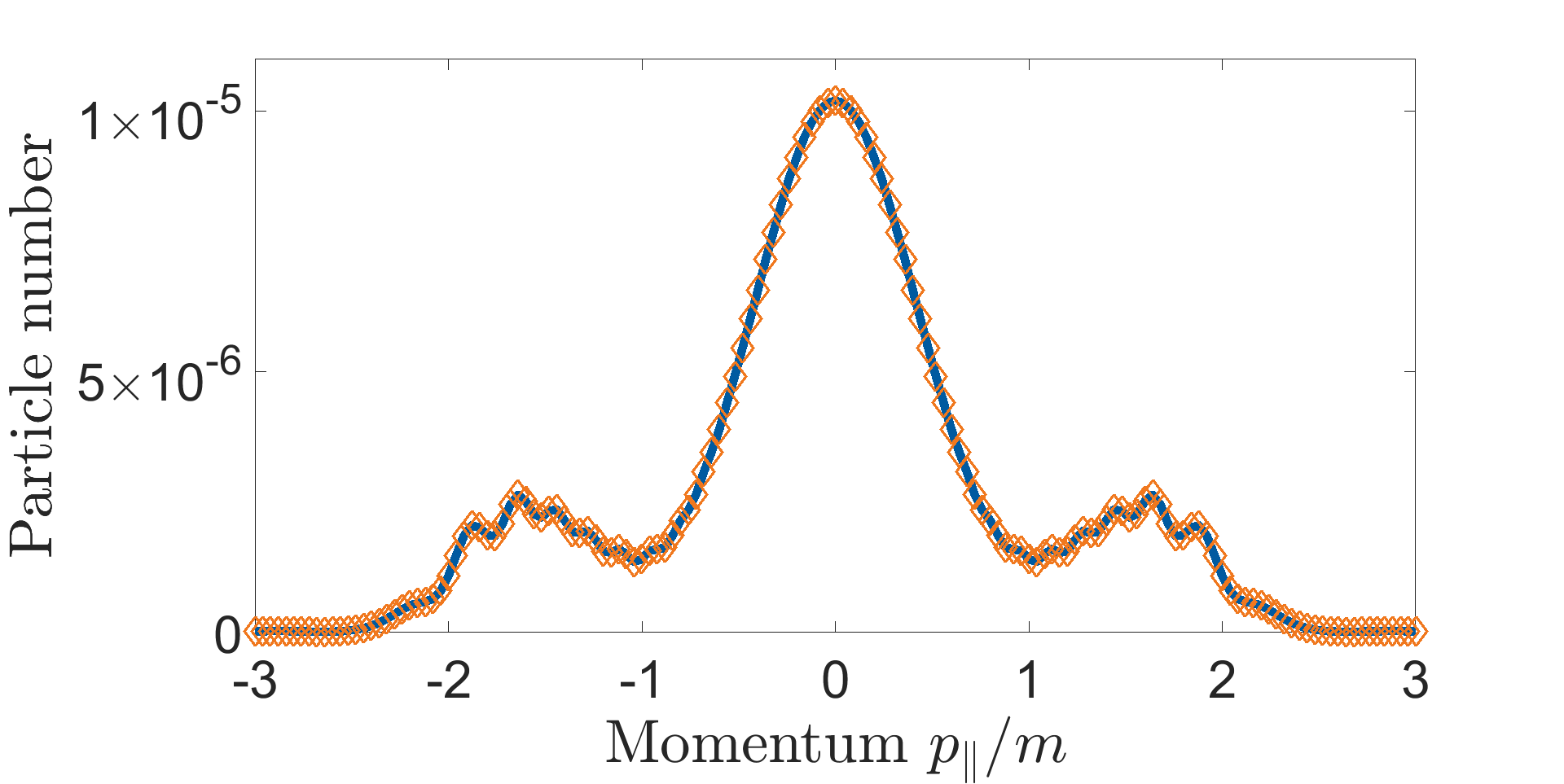}
 \caption{Asymptotic particle number as a function of the momentum $p_\parallel$ with $p_\perp=0$. The blue curve represents a calculation performed with the standard non-perturbative quantum kinetic equations. The orange markers are the result of a summation over contributions up to and including the sixth order in $\xi$.
 Field parameters: $\mathcal{E}=0.3,\ \varepsilon=0.05,\ \\\Omega=1.0m$,\ $\mathcal{T} = \tau = 20/m$.}
 \label{fig0}
\end{figure}

First and foremost, a convergence test of the derived hierarchical quantum kinetic equations is required. To do so, the results for a standard, non-perturbative calculation of the asymptotic particle number of created pairs are overlaid with the sum over all significant contributions in the alternative formulation, cf. Fig.~\ref{fig0}. %Quantitatively, the distributions functions integrated over parallel momentum are designated $n_{\rm stand.}$ and $n_{\rm pert.}$, respectively, to distinguish between standard and perturbative approaches. 
As such, we find for the scenario with parameters $\mathcal{E}=0.3,\ \varepsilon=0.05,\ \Omega=1.0m$,\ $\mathcal{T} = \tau = 20/m$ a deviation of roughly $10^{-8}$ \%. In essence, the results are in numerical agreement within limits of numerical precision.

For a pulse as defined in Eq.~\eqref{eq:A1}, the corresponding electric field has a single dominant peak in time, cf. Fig.~\ref{fig:A}. It is in the vicinity of this peak that the probability of particles emerging via the non-perturbative Schwinger effect is maximal. Consequently, a numerical calculation shows a smooth distribution function with a peak at vanishing momentum $p_\parallel$, cf. the blue curve in Fig.~\ref{fig1}a).

In a strongly oscillating, linearly polarized light field \eqref{eq:A2}, particle pairs are mainly produced by direct energy transfer from the light wave to the fermions. The energy carried by such a field is given by its oscillation frequency. For example, for a field of frequency $\Omega = m$, it is sufficient for the background field to provide two light quanta (with an energy of $m$ each) for the electron-positron pair to be produced, see Fig.~\ref{fig1}a), as the orange curve represents the perturbative result at order $\xi^4$.
Calculation of the particle spectra for the combined field $\mathcal{A}(t) + A^\gamma(t)$, cf. Fig.~\ref{fig1}b), then reveals the non-linear nature of strong-field pair production.

\clearpage

\begin{figure}[t]
%Onecolumn
 %\includegraphics[width=0.01\columnwidth]{fig1_run1both.png}
 %\put(0,100){a)}
 %\hfill
 %\includegraphics[width=0.43\columnwidth]{fig1_run1both.png}
 %\put(0,100){b)}
 %\hfill
 %\includegraphics[width=0.43\columnwidth]{fig2_run1both.png}
 %\hfill
 %\includegraphics[width=0.08\columnwidth]{legend_run1.png}
%Twocolumn
 \hspace{0.5cm}
 \includegraphics[width=0.66\columnwidth]{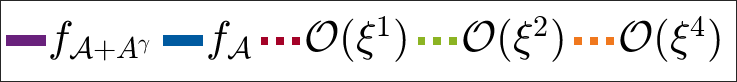} 
 \includegraphics[width=0.01\columnwidth]{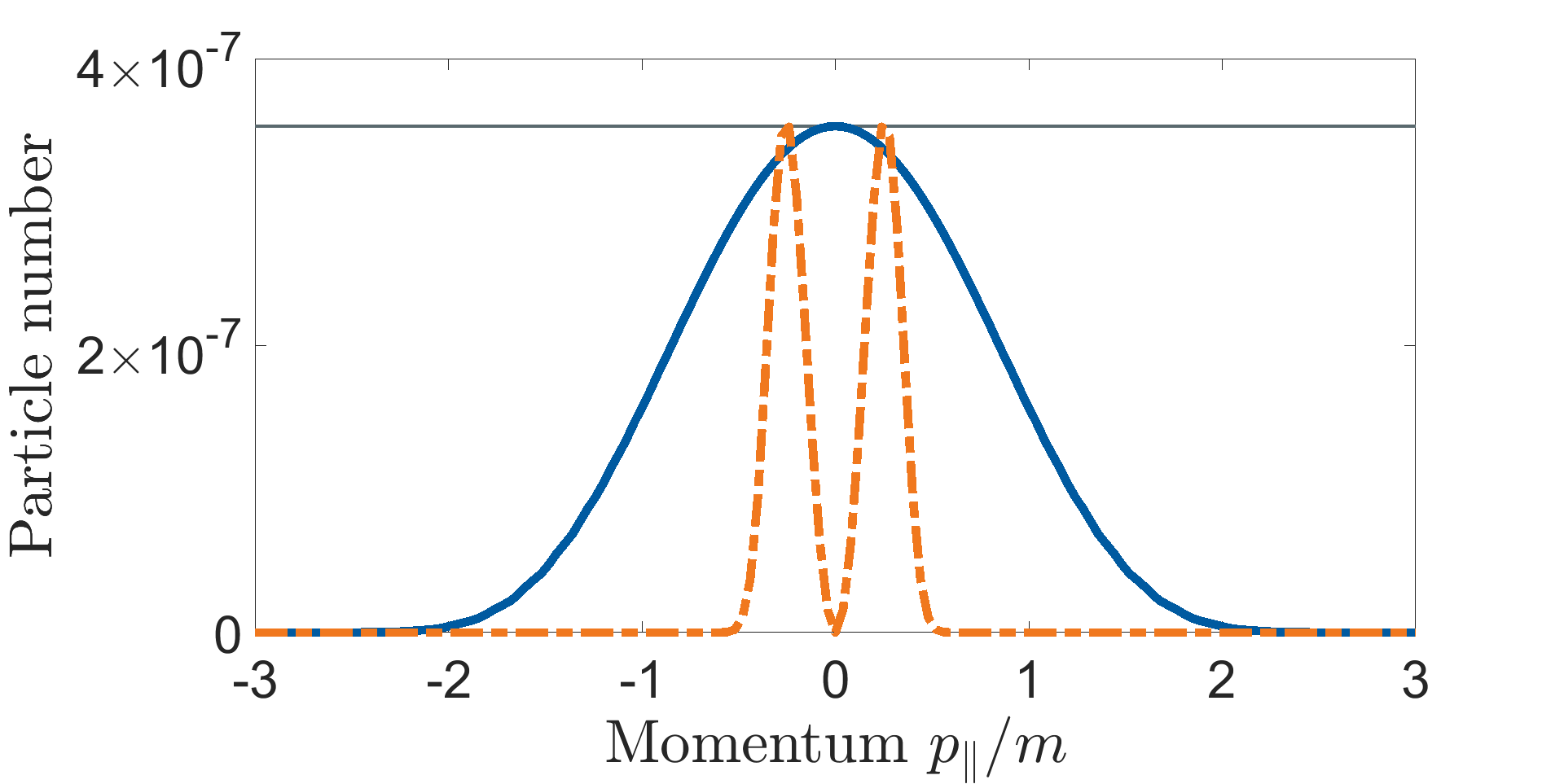}
 \put(-10,100){a)}
 \includegraphics[width=0.90\columnwidth]{fig1_run1both.png} 
 \\
 \includegraphics[width=0.01\columnwidth]{fig1_run1both.png}
 \put(-10,100){b)}
 \includegraphics[width=0.90\columnwidth]{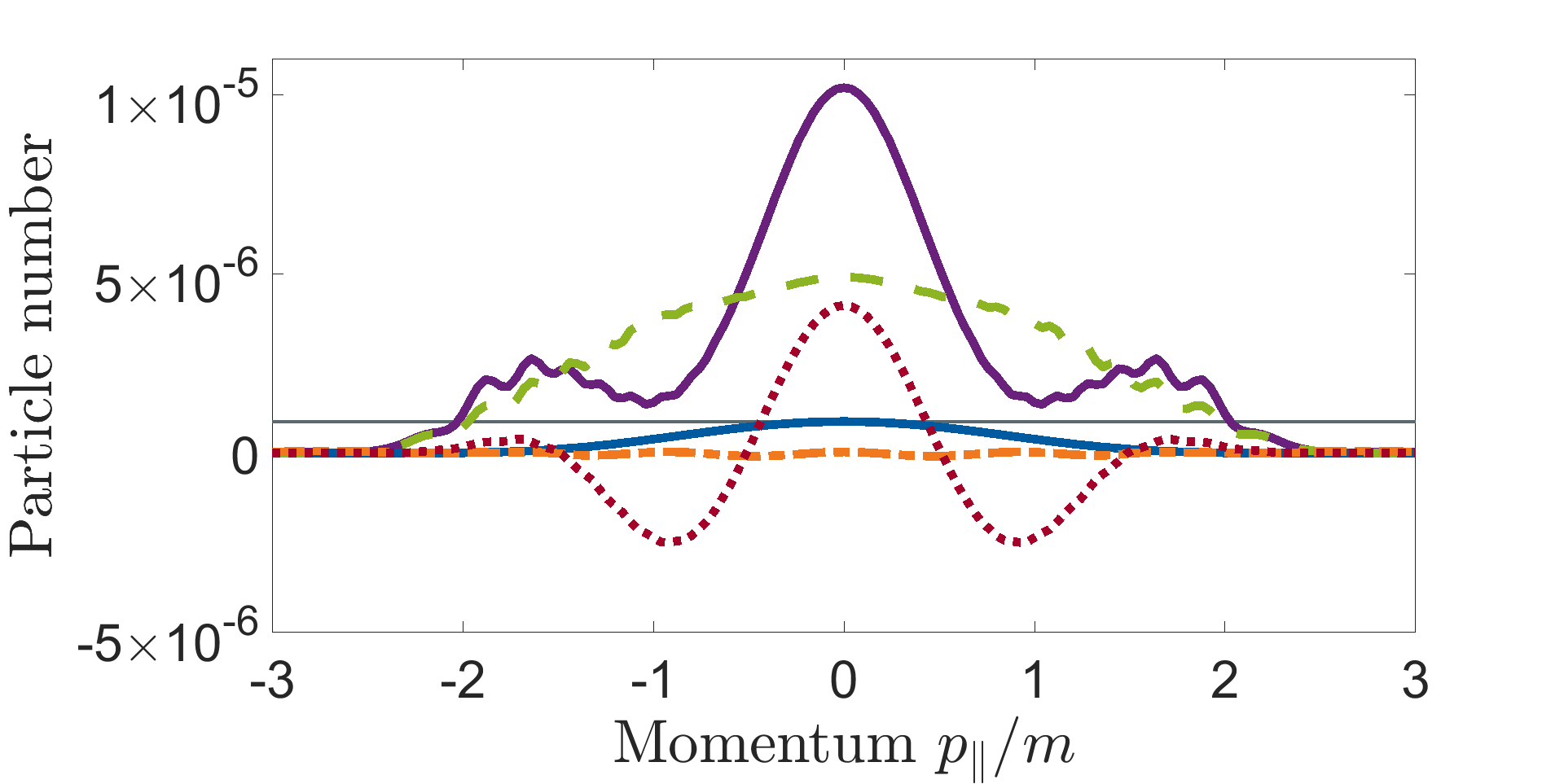}
 \caption{Spectral plot of the asymptotic particle number as a function of the momentum $p_\parallel$ with $p_\perp=0$. At the top, the two fields were used individually. The blue curve representing non-perturbative particle creation through the pulse only, while the orange curve corresponds to a calculation where only the oscillating field $A^\gamma(t)$ is non-zero. The latter corresponds to a two-quanta absorption signal.
 At the bottom, a field configuration with a slowly varying pulse superposed by an oscillating pulse was used. The dominant contributions to the full spectrum come from absorption of one energy quantum (green) and quantum interferences with the contributions from the pulse (red). The blue and orange curves correspond to the results for the slowly varying pulse and the perturbative contribution of order $\mathcal O(\xi^4)$, respectively. The latter includes the signal for absorbing two quanta of energy.
 Field parameters in use: $\mathcal{E}=0.3,\ \varepsilon=0.05,\ \Omega=1.0m$,\ $\mathcal{T} = \tau = 20/m$. The grey lines are placed at the same absolute height to help guide the eye.}
 \label{fig1}
\end{figure}

\begin{figure}[t]
 \centering
 \hspace{0.6cm}
 \includegraphics[width=0.725\columnwidth]{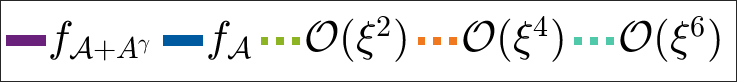}  
 \includegraphics[width=0.01\columnwidth]{legend7b.png}
 \put(-10,100){a)}
 \includegraphics[width=0.99\columnwidth]{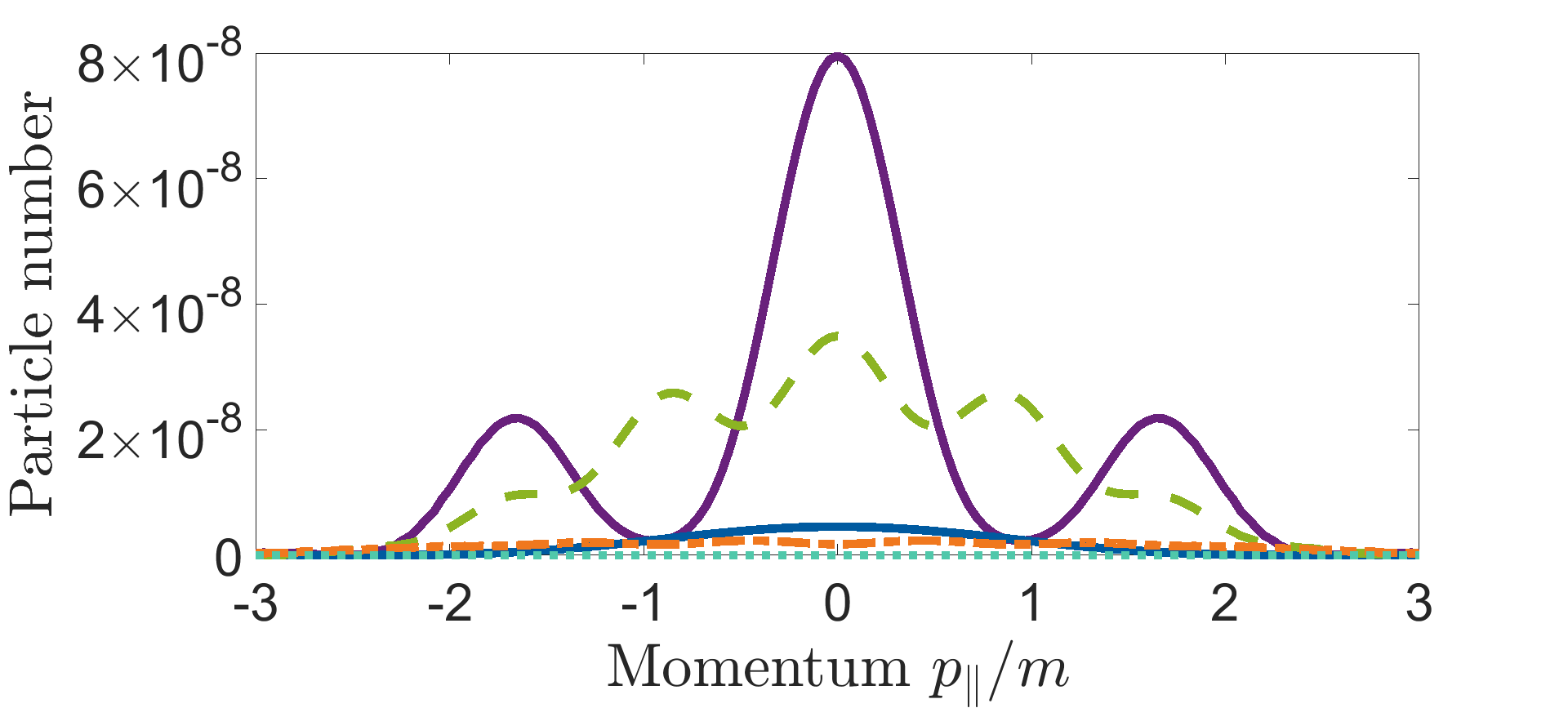}
 \includegraphics[width=0.01\columnwidth]{legend7b.png}
 \put(-10,100){b)}
 \includegraphics[width=0.99\columnwidth]{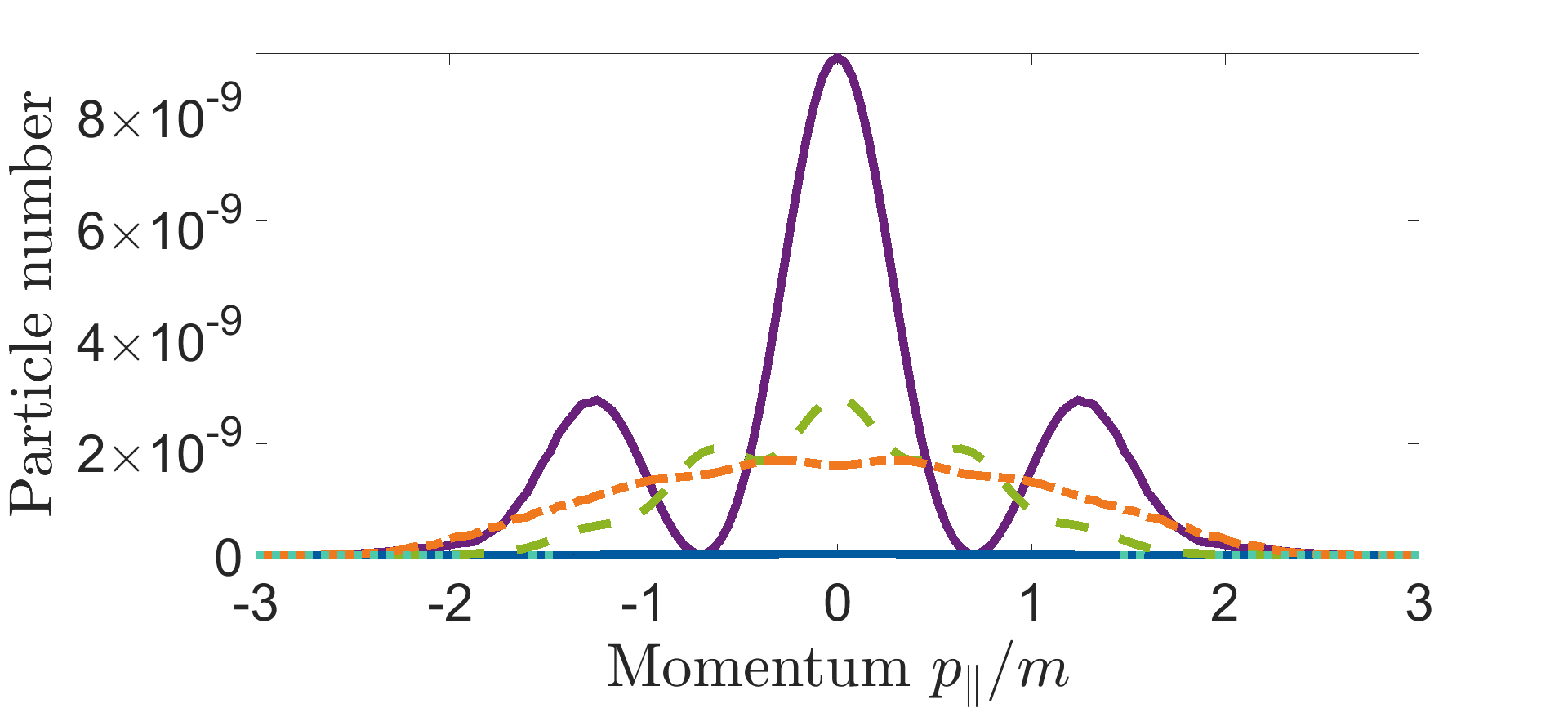}
 \includegraphics[width=0.01\columnwidth]{legend7b.png}
 \put(-10,100){c)}
 \includegraphics[width=0.99\columnwidth]{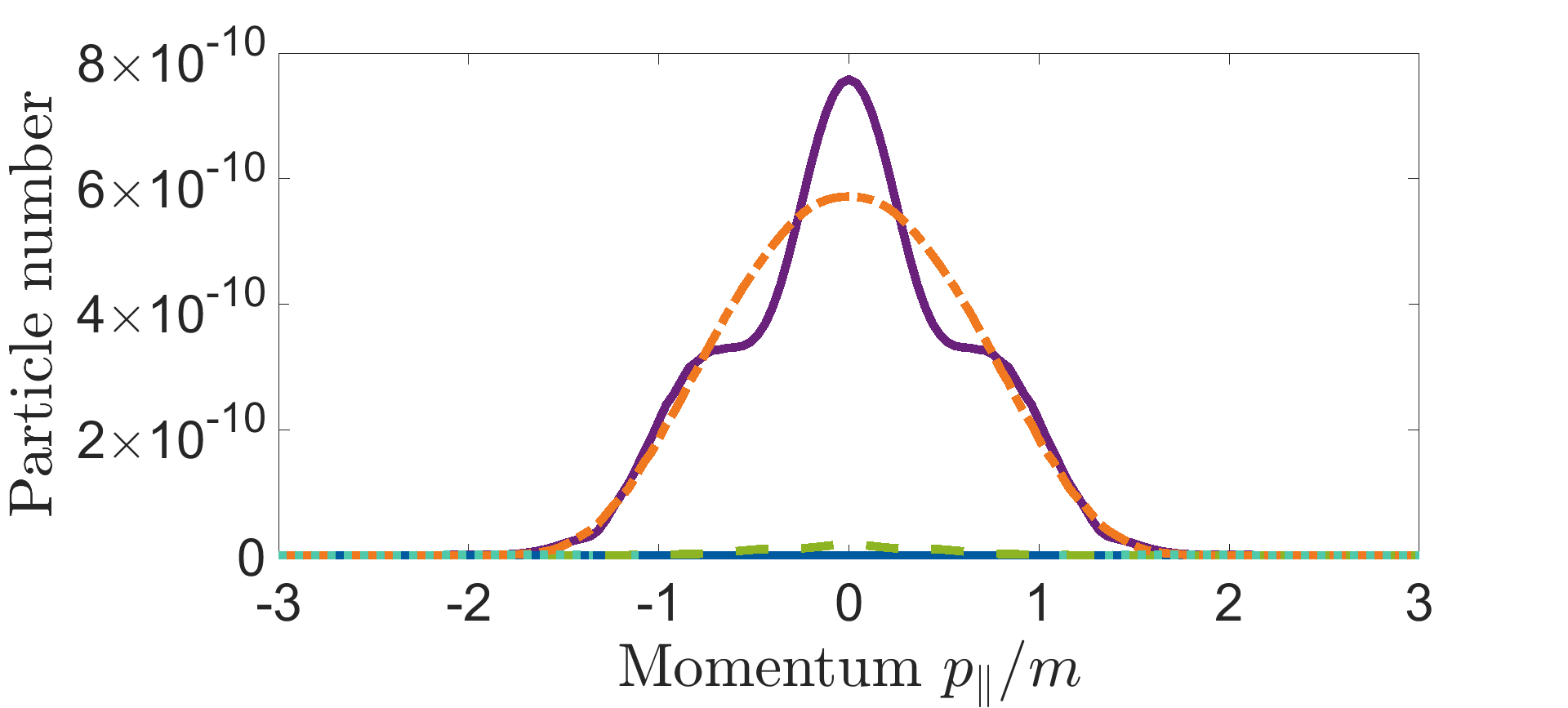}
 \includegraphics[width=0.01\columnwidth]{legend7b.png}
 \put(-10,100){d)}
 \includegraphics[width=0.99\columnwidth]{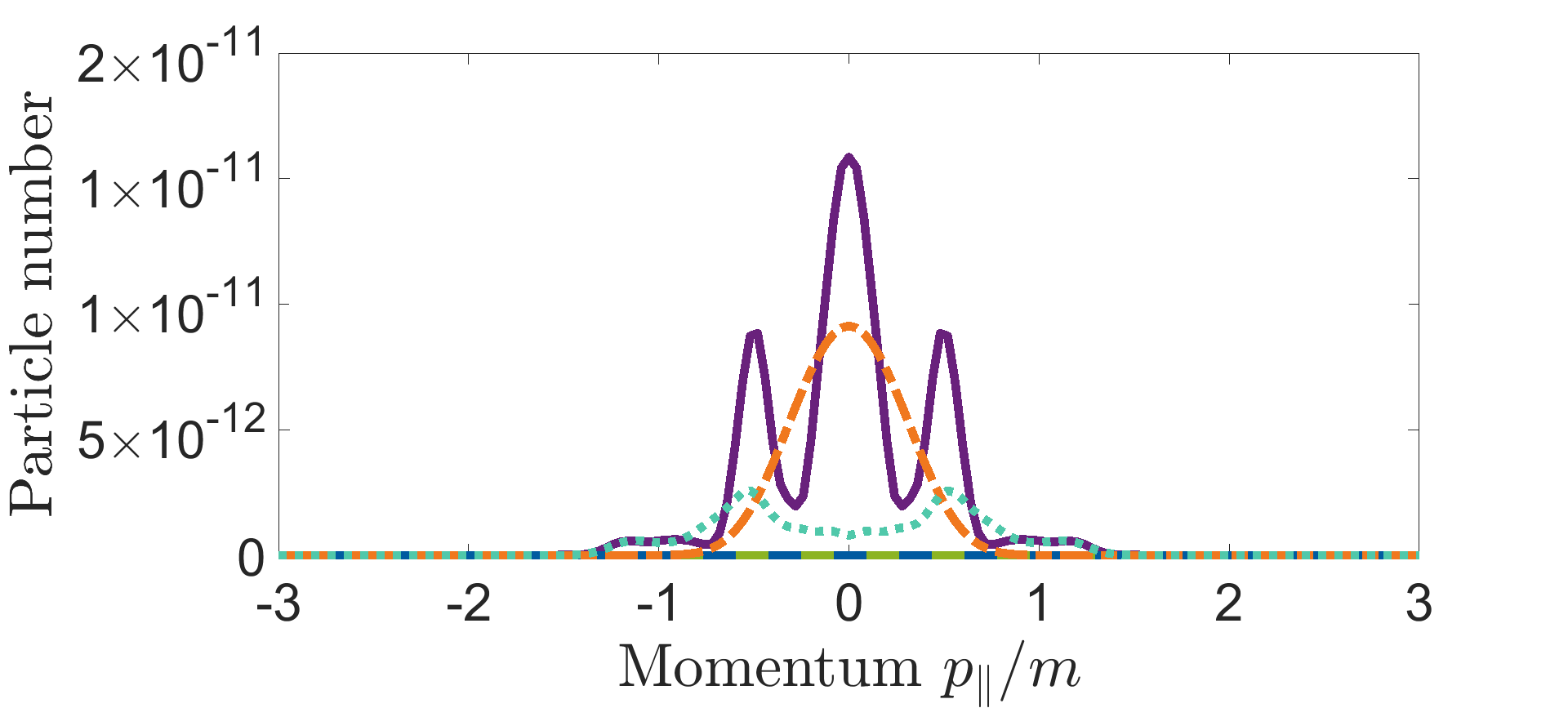}
 \caption{Series of plots of the asymptotic particle number as a function of momentum $p_\parallel$ with $p_\perp=0$ for both fields. The total particle yield is shown as the solid purple line. The contribution of the field $\mathcal A(t)$ to the total yield is shown in blue, and the effects of the orders $\mathcal O(\xi^2)$, $\mathcal O(\xi^4)$, and $\mathcal O(\xi^6)$, including the signals for absorbing one, two, and three quanta of energy, are shown in green, orange, and teal, respectively. From top to bottom, the field strength of the slowly varying field changes from $\mathcal{E}=0.2$ to $\mathcal{E}=0.15$, $\mathcal{E}=0.1$, $\mathcal{E}=0.05$. Other field parameters: $\varepsilon=0.02,\ \Omega=0.7m$,\ $\mathcal{T} = \tau = 40/m$. The total pulse duration of the configuration was extended to $40/m$ due to the lower field frequency $\Omega$ compared to the other results.}
 \label{fig3}
\end{figure}  

\clearpage

Figure \ref{fig3} presents a series of plots that provide additional context for the findings presented in the main article. Specifically, when all but the field strength of the slowly varying field $\mathcal E$ are held constant, an increase in yield is observed as $\mathcal E$ increases. Moreover, while assistance is observed throughout the series, the primary pathway for assistance undergoes a change. At the highest field strength, the original effect of one high-energy quanta boosting the particle production rate is recovered. Conversely, with only an oscillating field and $\Omega=0.7m$, there is only perturbative pair production by absorbing three quanta of energy. However, the transition between these two extremes is non-trivial, as with higher field strength, the two-quanta absorption channels first become dominant, and only at even higher field strengths does one-energy absorption become the dominant process. Note that, for the sake of clarity, the contributions in the odd orders of $\xi$ are not shown explicitly.

Finally, in Table~\ref{tab1} and Table~\ref{tab2}, we provide additional quantitative information on the numerical calculations.

\ctable[
label = tab1,
caption = {For a field with strength $\mathcal{E}=0.15$, cf. data of Fig.~\ref{fig3}b), the total yield is $\mathcal{N} = 9.12 \times 10^{-9}$. The yield is distributed between contributions from $f_{\mathcal{A}}$, cf. $\mathcal O (\xi^0)$, and higher orders in $\xi$.},
mincapwidth = \columnwidth,
]{lrclr}{
}{
    \toprule
    Contribution & [\%] & \hspace{2cm} & Contribution & [\%] \\
    \midrule
    %$f_{\mathcal{A}}$ & $0.47$ & & $\mathcal O (\xi^4)$ & $52.7$ \\
    $\mathcal O (\xi^0)$ & $0.47$ & & $\mathcal O (\xi^4)$ & $52.7$ \\
    $\mathcal O (\xi^1)$ & $0.16$ & & $\mathcal O (\xi^5)$ & $0.01$ \\
    $\mathcal O (\xi^2)$ & $46.8$ & & $\mathcal O (\xi^6)$ & $-0.14$ \\
    $\mathcal O (\xi^3)$ & $-0.06$ & & $\mathcal O (\xi^7)$ & $-0.002$ \\
    \bottomrule
}  

\ctable[
pos = t,
label = tab2,
caption = {Distribution functions $f$ evaluated at $p_\parallel = 0$ and $p_\parallel \approx \pm 0.72 m$, respectively, based on the data set of Fig.~\ref{fig3}b). The total particle distribution is divided into even, $\mathcal O (\xi^2) + \mathcal O (\xi^4)$, and odd, $\mathcal O (\xi^3)$, order contributions.},
mincapwidth = \columnwidth,
]{lrclr}{
}{
    \toprule
    f($p_\parallel = 0$) & $[10^{-9}]$ & \hspace{2cm} & f($p_\parallel = 0.72$) & $[10^{-9}]$ \\
    \midrule
    $f(p_\parallel)$ & $8.9$ & & $f(p_\parallel)$ & $0.02$ \\
    Even & $4.4$ & & Even & $3.3$ \\
    Odd & $4$ & & Odd & $-2.9$ \\
    \bottomrule
} 

%The corresponding distribution functions evaluated at $p_\parallel = 0$ and $p_\parallel \approx \pm 0.72 m$ at magnitude $10^{-9}$ are as follows: \nopagebreak
%\begin{itemize}
%\begin{minipage}{0.26\linewidth}
%    \item $f(p_\parallel)$
%    \item $\mathcal O (\xi^2) + \mathcal O (\xi^4)$
%    \item $\mathcal O (\xi^3)$
%\end{minipage}
%\begin{minipage}{0.15\linewidth}
%    \item[] $8.9 $
%    \item[] $4.4$
%    \item[] $4$
%\end{minipage}
%\begin{minipage}{0.26\linewidth}
%    \item $f(p_\parallel)$
%    \item $\mathcal O (\xi^2) + \mathcal O (\xi^4)$
%    \item $\mathcal O (\xi^3)$
%\end{minipage}
%\begin{minipage}{0.15\linewidth}
%    \item[] $0.02$
%    \item[] $+3.2$
%    \item[] $-2.9$
%\end{minipage}
%\end{itemize}  

%%%%%%%%%%%%%%%%%%%%%%%%%%%%%%%%%%%%%%%%%%%%%%%%%%%%%%%%%%%%%%%%%%%%%%%%%%%%%%%%%
%\paragraph{Bibliography} 

%\newpage 
\interlinepenalty=10000

%%%%%%%%%%%%%%%%%%%%%%%%%%%%%%%%%%%%%%%%%%%%%%%%%%%%%%%%%%%%%%%%%%%%%%%%%%%%%%%%%

\end{document}